\documentclass[12pt]{article}

\usepackage{cite}
\usepackage{amsmath}
\usepackage{amssymb}
\usepackage{indentfirst}
\usepackage{amssymb}
\usepackage{amsfonts}
\usepackage{amscd}
\usepackage{amsbsy}
\usepackage{amsthm}
\usepackage{latexsym}
\usepackage{graphicx,color} 
\usepackage[dvipsnames]{xcolor}
\usepackage[colorlinks]{hyperref}
\usepackage[normalem]{ulem} 
\hypersetup{linkcolor=blue,citecolor=blue,urlcolor=blue}
\font\mybb=msbm10 at 10pt
\def\bb#1{\hbox{\mybb#1}}
\def\be{\begin{equation}}
\def\ee{\end{equation}}
\def\bea{\begin{eqnarray}}
\def\eea{\end{eqnarray}}
\def\beq{\begin{eqnarray}}
\def\eeq{\end{eqnarray}}

\def\a{\alpha}\def\b{\beta}

\def\G{\Gamma}

\newcommand{\p}[1]{(\ref{#1})}

\numberwithin{equation}{section}

\begin{document}

\begin{center}

{\Large
Superembedding approach to superstrings and super-p-branes}\footnote{
Chapter for the Section ``String Theories" (Eds. C. Angelantonj and I. Antoniadis)
of the ``Handbook of Quantum Gravity" (Eds. C. Bambi, L. Modesto
and I.L. Shapiro, Springer Singapore, expected in 2023).}
\vskip 3mm

\textbf{Igor A. Bandos}$^{(a,b)}$\footnote{E-mail address: igor.bandos@ehu.eus}
\ \
and
\ \ \textbf{Dmitri P. Sorokin}$^{(c)}$\footnote{E-mail address:
dmitri.sorokin@pd.infn.it}

\vskip 4mm

(a) \ \ Department of Physics and EHU Quantum Center,
University of the Basque Country UPV/EHU,
\\
P.O. Box 644, 48080 Bilbao, Spain
\vskip 2mm

(b) \ \ IKERBASQUE, Basque Foundation for Science,
\\
48011, Bilbao, Spain
\vskip 2mm

(c) \ \ Istituto Nazionale di Fisica Nucleare, Sezione di Padova
\\
via F. Marzolo 8, 35135 Padova, Italia

\end{center}

%


\begin{quotation}

\noindent
\textbf{Abstract.}
We review a geometrical, so called superembedding, approach to the description of the dynamics of point-like and extended supersymmetric objects (superbranes) in String Theory. The approach is based on a supersymmetric extension of the classical surface theory to the description of superbrane dynamics by means of embedding worldvolume supersurfaces into target superspaces.
Lorentz harmonics, twistors and pure spinors are its intrinsic ingredients. Main new results obtained with this approach include the following ones. Being manifestly doubly supersymmetric (on the worldvolume and in target superspace) the superembedding approach explained that the local fermionic kappa-symmetry of the Green-Schwarz-like superbrane actions  originates from the conventional local supersymmetry of the worldvolume. It established or clarified a classical relationship between various formulations of the dynamics of superparticles and superstrings, such as the Neveu-Schwarz-Ramond and the Green-Schwarz formulation. The full set of the equations of motion of the M-theory five-brane was first derived with the use of this approach.
\vskip 2mm

\vskip 2mm


\end{quotation}

\tableofcontents

\newpage

\section{Introduction}
This Chapter is intended to give an introduction into a
geometrical, so called superembedding, approach to the description
of the dynamics of supersymmetric particles and extended
supersymmetric objects such as strings, membranes and higher-dimensional branes. These objects are part of string theory where they play an important role
in the perturbative and non-perturbative structure of the theory.

We will call these dynamical supersymmetric objects super-p-branes, where $p$ stands for the
dimensionality of the object, so that p=0 means a particle, p=1
means a string, p=2 means a membrane and so on. Among these there are
Dirichlet p-branes emerged as Dirichlet boundaries of open strings \cite{Sagnotti:1987tw,Horava:1989vt,Dai:1989ua,Horava:1989ga,Leigh:1989jq},
 an M2-brane \cite{Bergshoeff:1987cm,Bergshoeff:1987qx} and an M5-brane \cite{Duff:1990xz,Gueven:1992hh,Callan:1991ky,Gibbons:1993sv} which live in an eleven-dimensional
corner extending string theory to M-theory, and there are also more
exotic objects such as Neveu-Schwarz 5-branes, Kaluza-Klein
monopoles and some other (see \cite{Bergshoeff:1997gy,Bergshoeff:1997ak,Meessen:1998qm,Eyras:1998hn,Obers:1998fb,deBoer:2012ma}  for more details and references).

The worldvolume dynamics of all (or almost all) of these objects
in a space-time enlarged with Grassmann-odd spinorial directions
(i.e. in target superspace) is described by a well known
formulation which we shall conventionally call the Green-Schwarz
(GS) formulation constructed for superstrings in \cite{Green:1983wt,Green:1983sg}, though for different branes it has
been developed by different people. Let us particularly mention a 3-brane action in a $D=6$ superspace constructed by Hughes, Liu and Polchinski  \cite{Hughes:1986fa}  and the 11D supermembrane action of Bergshoeff, Sezgin and Townsend \cite{Bergshoeff:1987cm,Bergshoeff:1987qx}.

For the superstrings there exists an alternative formulation,
which was historically the first one, known under the name of
Neveu, Schwarz and Ramond, the NSR (fermionic or spinning) string
\cite{Neveu:1971rx,Ramond:1971gb,Deser:1976rb,Brink:1976sc}. Analogous formulation exists for spinning
particles \cite{Brink:1976sz,Gershun:1979fb}, while it seems not possible to construct
in the same way spinning membranes or higher dimensional branes
\cite{Howe:1977hp}.  \footnote{As
in the case of the `Green--Schwarz' (GS) formulation, we will use the alias `NSR' for the spinning particles and the spinning strings just for short.}

The GS and the NSR formulations are very well known, but there
are also less conventional formulations of superbrane dynamics
such as Lorentz-harmonic
formulations \cite{Sokatchev:1985tc,Sokatchev:1987nk,Nissimov:1987nv,Nissimov:1987ci,Bandos:1990ji,Galperin:1991gk,Delduc:1991ir,Bandos:1991bh,Bandos:1992np,Bandos:1992ze,Bandos:1992hu,Galperin:1992pz,Bandos:1993yc,Bandos:1994eu,Fedoruk:1994ij,Uvarov:2000wt,Uvarov:2008bx,Bandos:2007mi,Bandos:2007wm} and, in particular cases of superparticles and superstrings, twistor formulations
\cite{Ferber:1977qx,Shirafuji:1983zd,Bengtsson:1987ap,Eisenberg:1988nt,Plyushchay:1990eq,Chikalov:1992su,Chikalov:1993yv,Zima:1995db,Fedoruk:2005ks,Fedoruk:2006de}.

The Lorentz harmonic approach, especially in its form developed in \cite{Bandos:1990ji,Bandos:1991bh,Bandos:1992np,Bandos:1992hu,Bandos:1993yc,Bandos:1994eu}, which is also known under the name of
 spinor moving frame formalism, is actually closely related to the twistor approach. It provided a natural basis for a higher dimensional generalization of the twistor formulation \cite{Bandos:2006nr,Bandos:2007mi,Bandos:2007wm}. In addition to the 'standard' superparticles, superstrings and superbranes, in \cite{Bandos:2006af}
and \cite{Bandos:2014lja,Bandos:2019zqp} the spinor moving frame formalism was also applied to the twistor string
\cite{Witten:2003nn,Berkovits:2004hg}  and the ambitwistor string \cite{Mason:2013sva,Geyer:2014fka}. This allowed one to establish the relation of the four-dimensional twistor and ambitwistor strings to a so-called null-superstring (see \cite{Bandos:1990mw,Bandos:1993pa} and references therein) and to generalize the twistor string to higher dimensions \cite{Bandos:2006af} and the ambitwistor string to 11 (and 4) dimensions \cite{Bandos:2014lja}. The Lorentz harmonic  approach was also applied to the study of amplitudes of $D=10$ supersymmetric gauge theory and D=11 supergravity \cite{Bandos:2016tsm,Bandos:2017zap,Bandos:2017eof}.

The twistor and the Lorentz-harmonic formulations were proposed
as an attempt to solve the problem of the covariant quantization
of the superparticles and superstrings. But though
interesting and important results have been obtained using these methods (especially in the case of superparticles), the
problem of the covariant quantization of the GS superstring remained and became even
more topical in view of the interest in studying string
theory in non-trivial supergravity backgrounds with non-zero
antisymmetric gauge fields of the Ramond-Ramond type, such as Anti-de-Sitter (AdS) spaces,
pp--waves and string compactifications with Ramond-Ramon fluxes for which the
NSR formulation is not applicable.

The superembedding approach also arose from
the desire to solve the problem of the covariant quantization of
the Green--Schwarz superstring. The idea was to find a more
general (doubly-supersymmetric) formulation which would encompass main positive features
of the NSR, GS and twistor formulation, thus allowing one to make
a progress in quantizing the superstring covariantly. Such a
formulation was first realized for superparticles and then for
superstrings, and in the end for all known super-p-branes by efforts
of several theoretical groups \cite{Sorokin:1989zi,Volkov:1988vf,Sorokin:1988nj,Berkovits:1989zq,Berkovits:1990yr,Berkovits:1991qg,Tonin:1991ii,Tonin:1991ia,Delduc:1991if,Tonin:1992iq,Ivanov:1991ub,Gauntlett:1991dw,Townsend:1991sj,Pashnev:1992qy,Chikalov:1992su,Galperin:1992bw,Aoyama:1992fr,Delduc:1992fk,Berkovits:1993zy,Galperin:1993nb,Chikalov:1993yv,Bandos:1995zw,Howe:1996mx,Howe:1996yn,Howe:1997fb,Howe:1997wf,Howe:1998cp,Uvarov:2001tc,Uvarov:2001tz}.
In \cite{Bandos:1995zw} the spinor moving frame variables were incorporated into the doubly-supersymemtric approach and this gave rise to the present form of the superembedding formalism.

The progress in the covariant quantization of the superstring
during the last decades has been mainly made by
Berkovits \cite{Berkovits:1990yr,Berkovits:1991qg,Berkovits:1993zy,Tonin:1992iq}.
In 2000 he put forward a powerful quantization prescription \cite{Berkovits:2000fe,Berkovits:2000ph} based on the
use of pure spinors  (see e.g. \cite{Oda:2001zm,Oda:2004bg,Berkovits:2006vi,Oda:2007ak,Chandia:2007vp,Tonin:2010mm} for its developments
and \cite{Berkovits:2022fth} for the most recent review and a
detailed list of references). Geometrical roots of
these techniques stem from the superembedding approach
\cite{Matone:2002ft}. The relation of the pure spinor approach and the spinor moving frame approach in the cases of an $11D$ superparticle and a $10D$ superstring was discussed in
\cite{Bandos:2007mi,Bandos:2007wm} and \cite{Bandos:2012hp}.

The superembedding approach is an elegant, profound and in a sense universal geometrical
formulation, and the purpose of this Chapter is to explain its
main features (see \cite{Sorokin:1999jx,Bandos:2009xy} for detailed reviews).

The superembedding approach is based on a supersymmetric extension
of the classical surface theory and its application to bosonic relativistic strings \cite{Lund:1976ze,Omnes:1977km} to the description of superbrane
dynamics by means of embedding worldvolume supersurfaces into
target superspaces \cite{Bandos:1995zw}. Being manifestly doubly supersymmetric (on the
worldvolume and in target superspace) the superembedding approach
has explained the origin and the nature of a local fermionic symmetry
(so called $\kappa$-symmetry) of the GS formulation as a manifestation of the
conventional local supersymmetry of the worldvolume. In this way
it has solved the problem of infinite reducibility of the
$\kappa$-symmetry by realizing it as an irreducible extended
worldvolume supersymmetry \cite{Sorokin:1989zi}. As we have
mentioned, this stimulated progress in the covariant quantization
of the Green--Schwarz superstring.

The superembedding approach established or clarified a classical
relationship between various formulations of the dynamics of
superparticles and superstrings, such as the NSR and the GS
formulation \cite{Volkov:1988vf,Sorokin:1988nj,Townsend:1991sj,Aoyama:1992fr,Uvarov:2001tc}, the twistor 
and
harmonic 
descriptions.

This approach has proved to be a universal and powerful method
applicable to the description of all known supersymmetric branes,
in particular, to those of them for which standard methods
encountered problems because of their specific structure, such as
the 5--brane of M--theory. The superembedding methods allowed, for
the first time, to derive the complete set of covariant equations
of motion of the M5--brane \cite{Howe:1996yn,Howe:1997fb}. And only later these
equations were obtained \cite{Bandos:1997gm} from the M5-brane action
\cite{Bandos:1997ui,Aganagic:1997zq} based on a different technique adapted to deal
with duality-symmetric and self-dual (chiral) p-form fields \cite{Pasti:1995tn,Pasti:1996vs}. An alternative super-5-brane action  was proposed in \cite{Cederwall:1997gg}.
 However,  the self-duality of the  field strength of the 2-form gauge field on the M5-brane worldvolume does not follow from this action. Nevertheless, if imposed separately,  it is  consistent with the equations of motion of the model.

The superembedding formulation has also proved to be useful for
studying the ``brany'' mechanism of partial supersymmetry breaking
\cite{Hughes:1986dn,Hughes:1986fa,Achucarro:1988qb,Gauntlett:1990xq,Kallosh:1997aw} by giving a geometrical recipe
\cite{Pasti:2000zs,Bandos:2001ku,Drummond:2001uj} for the construction of
covariant worldvolume supersymmetric actions for superbranes. Upon
gauge fixing worldvolume superdiffeomorphisms these actions become
those of effective field theories with non--linearly realized
spontaneously broken supersymmetries. This has demonstrated an
intrinsic link of the superembedding approach and the method of
nonlinear realizations put forward in application to supersymmetric
theories in \cite{Volkov:1972jx,Volkov:1973ix,Volkov:1973jd,Volkov:1974ai,Bagger:1996wp,Rocek:1997hi,Ivanov:1999fwa,Bellucci:2000bd,Kapustnikov:2001za} (see \cite{Bellucci:2000kc,Ivanov:2016lha} for a
review and further references).

We now pass to a more detailed consideration of general properties
of this geometrical formulation starting from the description of its ingredients.

\section{Worldvolume vs target space supersymmetry}
\subsection{The `Neveu--Schwarz--Ramond' formulation}
In the NSR formulation which describes spinning particles
\cite{Brink:1976sz,Gershun:1979fb} and spinning strings~\cite{Neveu:1971rx,Ramond:1971gb,Deser:1976rb,Brink:1976sc}, as well as recently proposed ambitwistor string in its original formulation \cite{Mason:2013sva}, the worldline or
worldsheet of the spinning object is a supersurface ${\cal M}_{sw}$
parametrized by bosonic coordinates $\xi^m$ and fermionic
coordinates $\eta^\mu$ which we will collectively call
$z^M=(\xi^m,\eta^\mu)$.  Depending on the model considered ${\cal M}_{sw}$
may have a various number of fermionic directions $\eta$. The dynamics of the
spinning object is described by embedding ${\cal M}_{sw}$ into a
bosonic target spacetime $M_{TS}$ parametrized by coordinates
$X^{\underline m}$ ($\underline m =0,1,\cdots,D-1$).\footnote{To avoid the proliferation of the indices, in what follows we will denote by Latin indices the components of the bosonic directions and by Greek indices the components of fermionic directions of the worldvolume superspace. The underlined Latin and Greek indices are associated with target superspace directions, while (underlined) capital Latin indices stand for the both bosonic and fermionic directions.} In the
classical problems the number of space--time dimensions can be
arbitrary, but for quantum consistency the spinning string, whose
worldsheet has one or two fermionic directions, must live in a
ten--dimensional target space.

The motion of the spinning object is described by the image of
${\cal M}_{sw}$ in $M_{TS}$
\begin{equation}\label{soro-1}
X^{\underline m}(z^M)=x^{\underline m}(\xi)+i\eta\chi^{\underline
m}(\xi),
\end{equation}
where, for simplicity, we assumed that ${\cal M}_{sw}$ has the single Grassmann-odd direction parametrized by $\eta$, $x^{\underline m}(\xi)$ is associated with the bosonic
degrees of freedom of the spinning particle or string in ${M_{TS}}$,
and the Grassmann--odd vector $\chi^{\underline m}(\xi)$ is
associated with its spin degrees of freedom.

The NSR formulation is invariant under worldsheet
superdiffeomorphisms \linebreak $z^M~\rightarrow~z^{'M}(z^M)$, which include
 bosonic reparametrizations of ${\cal M}_w$
\begin{equation}\label{soro-2}
\delta\xi=a(\xi),
\end{equation}
and local worldsheet supersymmetry
\begin{equation}\label{soro-3}
\delta\eta=\kappa(\xi), \qquad \delta\xi=i\kappa(\xi)\eta.
\end{equation}

The presence of the local symmetries (\ref{soro-2}) and
(\ref{soro-3}) implies that the dynamics of the spinning objects
is subject to bosonic and fermionic first--class constraints,
respectively, which we will {\it schematically} write down in the
form
\begin{equation}\label{soro-4}
(\partial x^{\underline m})^2=0, \qquad \partial x^{\underline
m}\chi_{\underline m}=0.
\end{equation}
The bosonic constraint in (\ref{soro-4}) stands for the mass shell or the
Virasoro conditions, and the fermionic constraint produces, upon
quantization, the Dirac equation for the spin wave functions of
the dynamical system.

The NSR formulation does not have target--space supersymmetry. The
latter appears, in the case of the spinning string, only at the
quantum level upon imposing the Gliozzi--Scherk--Olive projection \cite{Gliozzi:1976qd}.
A merit of this formulation is that it is covariantly
quantizable \cite{Ito:1985qa}.

\subsection{The `Green--Schwarz' formulation}

This formulation is applicable to all known superparticles,
superstrings  and  conventional superbranes. Now the worldvolume ${\cal M}_w$ is
a (p+1)--dimensional bosonic surface parametrized by the
coordinates $\xi^m$ ($m=0,1,\cdots,p$) and the target space, into
which ${\cal M}_w$ is embedded, is a superspace $M_{TS}$
parametrized by bosonic coordinates $x^{\underline m}$
($\underline m =0,1,\cdots,D-1$) and by an appropriate number of
fermionic coordinates $\theta^{\underline\alpha}$
($\underline\alpha =1,\cdots,2n$)
\begin{equation}\label{soro-5}
Z^{\underline M}(\xi)=\left(x^{\underline
m}(\xi),\,\theta^{\underline\alpha}(\xi)\right).
\end{equation}
The GS formulation is manifestly invariant under bosonic reparametrizations \hbox{$\xi~ \rightarrow ~\xi'(\xi)$} of the worldvolume ${\cal
M}_w$ ,
and target space superdiffeomorphisms  $Z^{\underline
M}~\rightarrow~Z^{'\underline M}(Z^{\underline M})$, which in the
case of flat target superspace reduce to the translations along
$x^{\underline m}$ and to the global target--space supersymmetry
transformations
\begin{equation}\label{soro-6}
\delta\theta^{\underline\alpha}=\epsilon^{\underline\alpha},
\qquad \delta x^{\underline m}=i\bar\theta\Gamma^{\underline
m}\delta\theta.
\end{equation}
There is also another (non--manifest) local worldvolume fermionic
symmetry, so--called $\kappa$--symmetry, inherent to the GS
formulation. This is an important symmetry which implies and
reflects the existence of supersymmetric BPS brane--like solutions
of corresponding supergravity theories. It is thus responsible for
the `brane scan' (i.e. it prescribes which brane lives in which
target superspace). The $\kappa$--symmetry was first observed in
the case of superparticles \cite{deAzcarraga:1982dw,Siegel:1983hh} and then became an important ingredient of the Green-Schwarz-like actions for the superstring \cite{Green:1983wt,Green:1983sg}, the supermembrane \cite{Bergshoeff:1987cm,Bergshoeff:1987qx}, the Dp-branes \cite{Cederwall:1996uu,Cederwall:1996ri,Aganagic:1996pe,Bergshoeff:1996tu,Aganagic:1996nn} and the M5-brane \cite{Bandos:1997ui,Aganagic:1997zq}.  Kappa-symmetry transformations have the following
generic form (in flat target superspace):
\begin{equation}\label{soro-7}
\delta\theta^{\underline\alpha}=\Pi^{\underline\alpha}_{~~\underline\beta}
\kappa^{\underline\beta}(\xi), \qquad \delta x^{\underline
m}=-i\bar\theta\Gamma^{\underline m}\delta\theta,
\end{equation}
where $\Pi^{\underline\alpha}_{~~\underline\beta}$,  in the most of the cases is a half-rank projector
matrix,
whose form depends on the object under consideration.
 Because of the presence of the projector in the
$\kappa$--transformations only
half of $\kappa^{\underline\alpha}$, i.e. $n$ of the $2n$
Grassmann spinor components, effectively contribute to the
variation of the worldvolume fields. i.e. the $\kappa$-symmetry is (actually infinitely) reducible. However, without the use of some auxiliary fields, it is not possible to single out the independent components of the $\kappa$-symmetry transformations in a Lorenz covariant way. This causes the problem of the covariant quantization of the GS formulation.

To solve the problem of infinite reducibility of the
$\kappa$-symmetry one should try to find its irreducible
realization which is covariant in target superspace. A natural
assumption is that this should be an extended local worldvolume
supersymmetry (\ref{soro-3}) with the number of independent
parameters equal to the number of independent (irreducible)
$\kappa$-symmetries \cite{Sorokin:1989zi}. This reasoning brings us to the doubly-supersymmetric formulation.

\subsection{The doubly-supersymmetric formulation}

In this formulation the dynamics of superbranes is described by
embedding a worldvolume {\sl super}surface ${\cal M}_{sw}$
parametrized by the coordinates $z^M=(\xi^m,\eta^\alpha)$
($m=0,1,\cdots,p$), ($\alpha=1,\cdots, n)$ into a target {\sl
super}space $M_{TS}$ parametrized by the coordinates
$Z^{\underline M}=(X^{\underline m}, \Theta^{\underline\alpha})$
($\underline m=0,1,\cdots,D-1$), ($\underline\alpha=1,...,2n$).
Note that the number of the Grassmann directions of ${\cal
M}_{sw}$ is half the number of the Grassmann directions of
$M_{TS}$. Such a choice of the supermanifolds for superembedding
is caused by our desire to identify $n$ local supersymmetries on
${\cal M}_{sw}$ with $n$ independent $\kappa$-symmetries of the
GS formulation. But for some applications \cite{Volkov:1988vf,Sorokin:1988nj,Berkovits:1989zq,Tonin:1991ii,Tonin:1991ia,Aoyama:1992fr}, in
particular, as far as quantization is concerned
\cite{Berkovits:1991qg,Tonin:1992iq,Berkovits:1993zy,Matone:2002ft}, one may also consider an
embedding of supersurfaces ${\cal M}_{sw}$ with less number of
Grassmann directions. In these cases models can contain residual
non-manifest $\kappa$-symmetry.

Thus in the doubly supersymmetric formulation the degrees of
freedom of the superbranes are described by worldvolume
superfields
\begin{equation}\label{soro-8}
X^{\underline m}(z^M)=x^{\underline
m}(\xi)+i\eta^\alpha\chi^{\underline m}_\alpha(\xi)+~\cdots, \quad
\Theta^{\underline\alpha}(z^M)=\theta^{\underline\alpha}(\xi)
+\eta^\alpha\lambda^{\underline\alpha}_\alpha(\xi)+~\cdots,
\end{equation}
where $\cdots$ stand for the terms of higher order in
$\eta^\alpha$. These terms contain auxiliary fields and in
addition, for example in the case of the D--branes and the
M5--brane, include the gauge fields propagating on the
worldvolumes of these branes.

From eq. (\ref{soro-8}) we see that in the doubly supersymmetric
construction the number of degrees of freedom of the superbrane
roughly speaking doubles. We now have $x^{\underline m}(\xi)$
describing the bosonic oscillations of the brane, the Grassmann
`spin'--vectors $\chi^{\underline m}_\alpha(\xi)$ as in the NSR
formulation, the Green--Schwarz fermionic spinor degrees of
freedom $\theta^{\underline\alpha}(\xi)$ and their bosonic
counterparts $\lambda^{\underline\alpha}_\alpha(\xi)$. So if all
these worldvolume fields are independent the corresponding models
will not describe conventional superbranes. In this case one will
get, for instance, so called spinning superparticles and spinning
superstrings \cite{Gates:1985vk,KowalskiGlikman:1987pw,Sorokin:1989jj} which have more degrees of freedom
than the conventional NSR and GS dynamical systems. In addition to local worldvolume supersymmetry, they also have an infinite reducible $\kappa$-symmetry as an independent local fermionic symmetry.

To reach our goal of interpreting $\kappa$-symmetry as a
manifestation of the local worldvolume supersymmetry we should
find an appropriate doubly--supersymmetric description of the {\it
conventional} superbranes. For this we should impose constraints
on the superfields (\ref{soro-8}) which relate their components in
such a way that the independent physical degrees of freedom
described by these superfields will correspond to the standard GS
formulation. The geometrical meaning of these constraints is that
they cause the worldvolume supersurface ${\cal M}_{sw}$ to be
imbedded into the target superspace ${M}_{TS}$ in a specific way.

\subsection{The geometrical meaning of the superembedding condition}
The generic requirement for the superembedding to be
appropriate to the description of the dynamics of the superparticles, superstrings and super-p-branes is as follows. Let us consider a supersurface ${\cal M}_{sw}$ (to be associatied with the worldvolume of a super-p-brane), whose
geometry is described by supervielbein one--forms $e^a(z^M)$
($a=0,1,\cdots,p$) and $e^{\alpha}(z^M)$ ($\alpha=1,\cdots n$),
and a curved target superspace $M_{TS}$, whose geometry is
described by supervielbein one--forms $E^{\underline
a}(Z^{\underline M})$ ($\underline a=0,1,\cdots,D-1$) and
$E^{\underline\alpha}(Z^{\underline M})$
($\underline\alpha=1,\cdots 2n$)\footnote{Note that the
supergeometries of ${\cal M}_{sw}$ and $M_{TS}$ should be that of
corresponding supergravities, which implies that the torsions and
curvatures of ${\cal M}_{sw}$ and $M_{TS}$ are subject to
appropriate supergravity constraints.}.

The superembedding construction is manifestly covariant under superdiffeomorphism transformations of the superworldvolume coordinates
\begin{equation}\label{sdz}
z^M \to z'^{M}(z)\,.
\end{equation}
In the cases of superparticles and superstrings this local superworldvolume symmetry can be reduced to a suitable (superconformal) subgroup.

For the superembedding of ${\cal M}_{sw}$ into $M_{TS}$ to
describe a super-p-brane propagating in $M_{TS}$ the superwordvolume pullback
of $E^{\underline a}$ along the Grassmann directions of ${\cal M}_{sw}$ must
vanish, i.e. in
\begin{equation}\label{soro-21}
E^{\underline a}(Z(z))=e^{ a}E_a^{~\underline
a}+ e^{\alpha}E_\alpha^{~\underline a}
\end{equation}
the Grassmann components are zero
\begin{equation}\label{soro-22}
E_\alpha^{~\underline a}\left(Z(z)\right)=0.
\end{equation}
For instance, in the case of a flat target superspace (which, for simplicity we will mostly consider in this review) a supersymmetric vector supervielbein is given by a Volkov-Akulov one-form
\begin{equation}\label{VA1f}
E^{\underline a}=dX^{\underline a}-id\bar\Theta\Gamma^{\underline a}\Theta\,.
\end{equation}
Then the superembedding condition \eqref{soro-22} means that the  components of the pullback of \eqref{VA1f} along $e^\alpha$ vanish
\begin{equation}\label{flatse}
\nabla_\alpha X^{\underline a}-i\nabla_\alpha\bar\Theta\Gamma^{\underline a}\Theta=0\,,
\end{equation}
where $\nabla_\alpha$ are fermionic covariant derivatives in the (generically curved) superworldvolume ${\cal M}_{sw}$.

For the most of the superbranes (with some subtleties for the
space filling and codimension one branes
\cite{Akulov:1998bq,Howe:2000vk,Bandos:2001ku,Drummond:2001uj}), the superembedding
condition (\ref{soro-22}), accompanied by the $M_{TS}$ and/or
${\cal M}_{sw}$ supergravity constraints, implies that
\begin{itemize}
\item
the geometry of the superworldvolume ${\cal M}_{sw}$ is induced by
its embedding into $M_{TS}$, i.e. the ${\cal M}_{sw}$ supergravity
on the brane is not propagating;
\item
the dynamics of the superbrane is subject to the standard
constraints of the Green--Schwarz formulation, such as the
Virasoro constraints and their fermionic counterparts;
\item
$\kappa$--symmetry is a particular form of worldvolume
superdiffeomorphisms;
\item
the consistency of the superembedding condition results in the
same `brane scan' as that of the Green-Schwarz formulation.
\end{itemize}
In addition, when the number of the Grassmann directions of ${\cal
M}_{sw}$ is 16 or higher, the integrability of the superembedding
condition requires the worldvolume superfields to satisfy the
dynamical equations of motion of the superbrane \cite{Galperin:1993nb,Bandos:1995zw,Howe:1996mx}
\footnote{This was a reason to the name `geometrodynamic equation' used for the superembedding condition  in  \cite{Delduc:1992fk,Bandos:1995zw}. Note also that this is similar to the case of maximally supersymmetric
super-Yang-Mills and supergravity theories in $D\geq 3$ whose superfield
constraints produce the dynamical equations of motion. }. It is in
this way that the covariant equations of motion of the M5--brane were
obtained for the first time \cite{Howe:1996yn,Howe:1997fb}. We will review the superembedding description of the M2- and M5-brane in Section \ref{25}.

\section{Superembedding description of superparticles}
\subsection{$N=1$, $D=3$ superparticle}
Let us now consider the  dynamical consequences of the superembedding condition in the simplest case of a superparticle propagating
in a flat $N=1$, $D=3$ target superspace \cite{Sorokin:1989zi}. Then the supersurface
${\cal M}_{sw}$ of the previous subsection is associated with the
superparticle ``worldline" having one bosonic (time) and one
fermionic coordinate $(\xi,\eta)$, and the target superspace is
parametrized by bosonic three--vector coordinates $X^{\underline
m}$ (${\underline m}=0,1,2$) and Grassmann Majorana two--spinor
coordinates $\Theta^{\underline\alpha}$ $(\underline\alpha=1,2)$.
The superembedding condition \eqref{flatse} relates the worldline superfields
$X^{\underline m}(z^M)$ and $\Theta^{\underline\alpha}(z^M)$ in
the following way (for simplicity we assume the geometry of the target superspace be flat)
\begin{equation}\label{soro-9}
DX^{\underline m}-iD\bar\Theta\Gamma^{\underline m}\Theta=0,
\end{equation}
where $D$ is a Grassmann covariant derivative on ${\cal M}_{sw}$
which in the case of the superparticles can be chosen to be flat
\begin{equation}\label{soro-10}
D={\partial\over{\partial\eta}}+i\eta{\partial\over{\partial\xi}},
\qquad \{D,D\}=2i\partial_\xi.
\end{equation}
Using the $\eta$--expansion (\ref{soro-8}), which in the case under
consideration does not contain ``$\cdots$"--terms, we obtain the
following relation between the components of the superfields
$X^{\underline m}(z^M)$ and $\Theta^{\underline\alpha}(z^M)$
\begin{equation}\label{soro-11}
\partial_\xi x^{\underline
m}-i\partial_\xi\bar\theta\Gamma^{\underline
m}\theta=\bar\lambda\Gamma^{\underline m}\lambda,
\end{equation}
\begin{equation}\label{soro-12}
\chi^{\underline m}=\bar\theta\Gamma^{\underline m}\lambda.
\end{equation}
From eq. (\ref{soro-11}) we see that
$\lambda^{\underline\alpha}(\xi)$ are not independent fields but
are expressed in terms of the derivatives of $x^{\underline m}$
and $\theta^{\underline\alpha}$. Moreover in the l.h.s. of
(\ref{soro-11}) one can recognize the canonical momentum of the
superparticle $P^{\underline m}={1\over {e(\tau)}}(\partial_\xi
x^{\underline m}-i\partial_\xi\bar\theta\Gamma^{\underline
m}\theta)$ \footnote{$e(\xi)$ is a proportionality coefficient.
Its geometrical meaning is to be the worldline gravitational
field.} whose square is identically zero
\begin{equation}\label{soro-13}
P^{\underline m}P_{\underline m}=0,
\end{equation}
 due to its so called Cartan--Penrose (or twistor)
representation as a bilinear combination of commuting (twistor-like) spinor
components $\lambda$ and the $\Gamma$--matrix identities
\begin{equation}\label{gammafierz}
\Gamma^{\underline
m}_{\underline\alpha\underline\beta}\Gamma_{\underline m
\underline\gamma\underline\delta}+\Gamma^{\underline
m}_{\underline\beta\underline\gamma}\Gamma_{\underline m
\underline\alpha\underline\delta}+\Gamma^{\underline
m}_{\underline\gamma\underline\alpha}\Gamma_{\underline m
\underline\beta\underline\delta}=0.
\end{equation}
We conclude that the superparticle is massless.

{\it In the case of the superstrings the superembedding condition
will produce in a similar way the Virasoro constraints, and it
will produce the corresponding constraints for the other
superbranes.}

Eq. (\ref{soro-12}) implies the relation between the Grassmann
vector and the Grassmann spinor variables, so that only one or
another can be taken to describe independent fermionic degrees of
freedom. This is a basic relation which allows one to establish a
classical correspondence between the NSR and the GS formulation of
supersymmetric particles and strings \cite{Volkov:1988vf,Sorokin:1988nj,Townsend:1991sj,Aoyama:1992fr,Uvarov:2001tc}.

\subsection{Local worldvolume supersymmetry versus
$\kappa$--symmetry}\label{ksusy}

Let us now demonstrate how $\kappa$--symmetry appears in the
superembedding formulation as a weird realization of the local
worldvolume supersymmetry.

The components (\ref{soro-7}) of the superfields $X^{\underline
m}(z^M)$ and $\Theta^{\underline\alpha}(z^M)$ transform under the
local worldline supersymmetry (\ref{soro-3}) in the standard way
\begin{equation}\label{soro-14}
\delta\theta^{\underline\alpha}=-\lambda^{\underline\alpha}\kappa(\xi),
\qquad\delta
\lambda^{\underline\alpha}=i\partial_\xi\theta^{\underline\alpha}\kappa(\xi),
\end{equation}
\begin{equation}\label{soro-15}
\delta x^{\underline m}=i\chi^{\underline m}\kappa(\xi), \quad
\qquad \delta\chi^{\underline m}=-\partial_\xi x^{\underline
m}\kappa (\xi).
\end{equation}
We now substitute into the first equation of (\ref{soro-15}) the
solution (\ref{soro-12}) of the superembedding condition
(\ref{soro-9}) and observe that, due to the form of the
$\theta$--variation (\ref{soro-14}), the variation of
$x^{\underline m}$ can be rewritten as follows
\begin{equation}\label{soro-16}
\delta x^{\underline m}=i(\bar\theta\Gamma^{\underline
m}\lambda)\,\kappa(\xi)=-i\bar\theta\Gamma^{\underline
m}\delta\theta.
\end{equation}

The next step is to replace the Grassmann scalar parameter of the
local supersymmetry by the scalar product of the spinor
$\lambda_{\underline\beta}$ with a Grassmann spinor parameter
$\kappa^{\underline\beta}(\xi)$, which is always possible,
\begin{equation}\label{soro-17}
\kappa(\xi)=2\lambda_{\underline\beta}\kappa^{\underline\beta}(\xi),
\end{equation}
and to substitute (\ref{soro-17}) into the $\theta$--variation
(\ref{soro-14}). We thus get
\begin{equation}\label{soro-18}
\delta\theta^{\underline\alpha}=
-2\lambda^{\underline\alpha}\lambda_{\underline\beta}\kappa^{\underline\beta}(\xi).
\end{equation}
We now note that, due to the superembedding condition
(\ref{soro-11}), the bilinear combination of $\lambda$ in
(\ref{soro-18}) is nothing but
\begin{equation}\label{soro-19}
\Pi^{\underline{\alpha}}_{~~\underline\beta}=
-2\lambda^{\underline\alpha}\lambda_{\underline\beta}=(\partial_\xi
x^{\underline m}-i\partial_\xi\bar\theta\Gamma^{\underline
m}\theta)(\Gamma_{\underline
m})^{\underline\alpha}_{~~\underline\beta},
\end{equation}
which is the projector matrix in the $\kappa$--symmetry variation
of $\theta$ (\ref{soro-7}). Hence, the local supersymmetry
variations (\ref{soro-16}) and (\ref{soro-18}) reduce to the
$\kappa$--variations (\ref{soro-7}).

{\sl We have thus demonstrated how, in virtue of the
superembedding condition (\ref{soro-9}), the $\kappa$--symmetry of
the GS formulation of the superbranes arises
 from the irreducible
local worldvolume supersymmetry.}

One might have already noticed the difference in sign in the
target--space supersymmetry variations of $x^{\underline m}$
(\ref{soro-6}) and in the worldvolume supersymmetry variations
(\ref{soro-16}) and corresponding $\kappa$--variations
(\ref{soro-7}). The rigid target--space supersymmetry and the local
worldvolume supersymmetry (or $\kappa$--symmetry) can be therefore
regarded as, respectively, `left' and (a part of) `right' supertranslations of
$x^{\underline m}$.

\subsection{The worldline superfield action for the $N=1$, $D=3$ superparticle}\label{D3}
The worldline $n=1$ superfield action, which produces the
superembedding relations \p{soro-9}, \p{soro-11} and \p{soro-12},
and also the superparticle equations of motion, has the following
form
\begin{equation}\label{n1d3}
S= -i\int d\xi d\eta P_{\underline m}\left(D X^{\underline m}-
iD\bar\Theta\Gamma^{\underline m}\Theta\right)\,,
\end{equation}
where $X^{\underline m}(\xi,\eta)$ and
$\Theta^{\underline\alpha}(\xi,\eta)$ are the superfields defined in
\p{soro-8} and
\begin{equation}\label{P}
 P_{\underline m}(\xi,\eta)=p_{\underline
m}(\xi)+i\eta\rho_{\underline m}(\xi)
\end{equation}
is the Lagrange multiplier superfield.

Varying (\ref{n1d3}) with respect to $P_{\underline m}$ we get the
superembedding condition (\ref{soro-9}), and hence \p{soro-11} and
\p{soro-12}.

To see what kind of other equations follow from the action let us
integrate it over $\eta$ using the Berezin rules $\int d\eta = 0$,
$\int d\eta \eta= 1$. As a result we get the following action for
the components of the superfields (\ref{soro-8}) and (\ref{P})
\begin{equation}\label{2.5.3}
S=\int d\xi \,p_{\underline m}(\partial_\xi x^{\underline m}-
i\partial_\xi{\bar\theta}\Gamma^{\underline m}\theta-\bar\lambda
\Gamma^{\underline m}\lambda)+ i\int d\xi\,\rho_{\underline
m}(\chi^{\underline m}- \bar\lambda\Gamma^{\underline m}\theta).
\end{equation}
We see that $p_{\underline m}$ is the particle momentum, and that
the second term in (\ref{2.5.3}) means that $\rho_{\underline a}$
and $\chi^{\underline m}$ are auxiliary fields satisfying
algebraic equations
\begin{equation}\label{2.5.4}
\rho_{\underline m}=0, \qquad \chi^{\underline m}=
\bar\lambda\Gamma^{\underline m}\theta,
\end{equation}
the latter being just part of the superembedding condition.
Another part of the superembedding condition, namely eq. \p{soro-11}, is obtained from \p{2.5.3} as a variation of $p_{\underline m}$.

The equations of motion of $x^{\underline m}$ and
$\theta^{\underline\alpha}$ are, respectively,
\begin{equation}\label{2.5.5}
\partial_\xi p_{\underline m}=0 \quad {\rm and} \quad
(p_{\underline m}\Gamma^{\underline m})^{\underline\alpha}
_{~~\underline\beta}\partial_\xi\theta^{\underline\beta}=0.
\end{equation}
These are the standard equations of motion of a
superparticle \cite{Siegel:1983hh}. What remains to show is that the particle under
consideration is massless, i.e. that its momentum is light-like
$p^{\underline m}p_{\underline m}=0$. To see this one should
consider the equation of motion of the commuting spinor variable
$\lambda^{\underline\alpha}$:
\begin{equation}\label{l}
(p_{\underline m}\Gamma^{\underline m})^{\underline\alpha}
_{~~\underline\beta}\,\lambda^{\underline\beta}=0.
\end{equation}
The general solution of this equation in $D=3$, as well as in
$D=4,6$ and $10$ is
\begin{equation}\label{pll}
 p_{\underline m}=e(\xi)\bar\lambda\Gamma_{\underline m}\lambda=
e(\xi)(\partial_\xi x^{\underline
m}-i\partial_\xi\bar\theta\Gamma^{\underline m}\theta)\,,
\end{equation}
where $e(\xi)$ is an arbitrary worldline function, and the last
equality being the consequence of \p{soro-11}. Hence
$p^{\underline m}p_{\underline m}=0$ in virtue of the Dirac matrix
identity \eqref{gammafierz}, and the particle is indeed
massless.

On the other hand, \eqref{pll} implies that
\begin{equation}\label{llp}
\bar\lambda\Gamma_{\underline m}\lambda=\frac {p_{\underline m}}{ e{\scriptstyle(\xi)}}.
\end{equation}
So, if we substitute eqs. \eqref{2.5.4} and the expression \eqref{llp} for the bilinear of  $\lambda$ into the action \eqref{2.5.3}, it reduces to a standard first-order action for a massless superparticle
\begin{equation}\label{2.5.31}
S=\int d\xi \left(\,p_{\underline m}(\partial_\xi x^{\underline m}-
i\partial_\xi {\bar\theta}\Gamma^{\underline m}\theta)-{e^{-1}{\scriptstyle(\xi)}}\,p_{\underline m}p^{\underline m}\right).
\end{equation}

Let us now consider constraints on the canonical fermionic momenta
$\pi_{\underline\a}={{\delta
S}\over{\delta\dot\theta^{\underline\a}}}$ conjugate to
$\theta^{\underline\a}$. They are
\begin{equation}\label{Da}
D_{\underline\a}=\pi_{\underline\a}-ip_{\underline
m}(\Gamma^{\underline m}\theta)_{\underline\a}=0\,.
\end{equation}
The Poisson brackets of these constraints are
\begin{equation}\label{pb}
\{D_{\underline\a},D_{\underline\b}\}=-2ip_{\underline
m}\Gamma^{\underline m}_{\underline\a\underline\b}\,.
\end{equation}
Since the momentum $p_{\underline m}$ is constrained \p{pll} to be
light--like, the r.h.s. of \p{pb} is a degenerate $2\times 2$
matrix of rank one. This means that the fermionic constraints
\p{Da} are the mixture of a first and a second class constraint.
In the conventional Green--Schwarz formulation it is not possible
to separate in \p{Da} an {\it irreducible} set of the first class
fermionic constraints from the second class ones in a
Lorentz--covariant way, which causes the covariant quantization
problem. The availability of auxiliary commuting spinor variables
$\lambda^{\underline\a}$ in the superembedding formulation allows
one to do so. One can convince oneself that the constraint
$\lambda^{\underline\a} D_{\underline\a}$ obtained by projecting
$D_{\underline\a}$ along $\lambda^{\underline\a}$ is the first
class
\begin{equation}\label{lD}
\{\lambda^{\underline\a} D_{\underline\a},\lambda^{\underline\b}
D_{\underline\b}\}=-2ip_{\underline m}\lambda\G^{\underline
m}\lambda \simeq0\,.
\end{equation}
The r.h.s. of \p{lD} is weakly zero in the Dirac sense due to
\p{l}.

The first class bosonic constraint $p_{\underline
m}\lambda\G^{\underline m}\lambda$ and the first class fermionic
constraint $\lambda^{\underline\a} D_{\underline\a}$  form the
one--dimensional (worldline) $n=1$ superalgebra and generate local
worldline reparametrizations and supersymmetry transformations,
the latter being the {\it irreducible} counterpart of the
kappa--symmetry of the Green--Schwarz formulation, as we have
discussed in Subsection 3.1. This provides us with an algebraic
ground for the covariant quantization of superparticles and
superstrings. Actually, as was realized in
\cite{Berkovits:1990yr,Berkovits:1991qg,Berkovits:1993zy,Tonin:1992iq}, a version of the superembedding
formulation with manifest $n=2$ worldsheet supersymmetry turns out
to be the most appropriate for quantizing $D=10$ superparticles
and superstrings. It is this version, which gives rise to pure
spinors and is in the origin of the covariant quantization
procedure of \cite{Berkovits:2022fth}. So we shall now
turn to the consideration of an $N=1$, $D=10$ superparticle with $n=2$ worldline
supersymmetry.

\subsection{The $N=1$, $D=10$ superparticle with $n=2$ worldline
supersymmetry and pure spinors}\label{D10pure}

We are now in a ten--dimensional flat superspace parametrized by
ten bosonic coordinates $X^{\underline m}$ ($\underline
m=0,1,\cdots,9$) and by sixteen real Majorana--Weyl Grassmann
coordinates $\Theta^{\underline\a}$ $(\underline\a=1,\cdots,16)$.

The superparticle worldline is assumed to be a supersurface ${\cal
M}_{sw}$ with one bosonic and two fermionic directions
$(\xi,\eta,\bar\eta)$, where $\eta$ and $\bar\eta$ are complex
conjugate to each other. The worldline fermionic covariant
derivatives
$$
D={\partial\over{\partial\eta}}+i\bar\eta{\partial\over{\partial\xi}},\qquad
\bar
D={\partial\over{\partial\bar\eta}}+i\eta{\partial\over{\partial\xi}},\qquad
$$
generate the worldline $n=2$ superalgebra
\begin{equation}\label{DbarD}
D^2=0,\qquad {\bar D}^2=0, \qquad \{D,\bar
D\}=2i{\partial\over{\partial\xi}}\,.
\end{equation}
The $n=2$ superfield form of the $N=1$, $D=10$ superparticle
action is
\begin{equation}\label{n2d10}
S= -i\int d\xi d\eta d\bar\eta\left[ P_{\underline m}\left(D
X^{\underline m}- iD\Theta\Gamma^{\underline m}\Theta\right)+{\bar
P}_{\underline m}\left(\bar D X^{\underline m}- i\bar
D\Theta\Gamma^{\underline m}\Theta\right)\right]\,,
\end{equation}
where $P_{\underline m}(\xi,\eta,\bar\eta)$ and $\bar
P_{\underline m}(\xi,\eta,\bar\eta)$ are complex conjugate
Lagrange multipliers, whose variation produces the superembedding
conditions
\begin{equation}\label{se2}
D X^{\underline m}- iD\Theta\Gamma^{\underline a}\Theta=0, \qquad
\bar D X^{\underline m}- i\bar D\Theta\Gamma^{\underline
m}\Theta=0
\end{equation}
to be satisfied by the superfields
$$
X^{\underline m}(\xi,\eta,\bar\eta)=x^{\underline
m}(\xi)+i\eta\chi^{\underline
m}(\xi)+i\bar\eta\bar\chi^{\underline m}(\xi)+\eta\bar\eta
y^{\underline m}(\xi)\,,
$$
$$
\Theta^{\underline\a}(\xi,\eta,\bar\eta)=\theta^{\underline\a}(\xi)+
\eta\lambda^{\underline\a}(\xi)+\bar\eta\bar\lambda^{\underline\a}(\xi)+
\eta\bar\eta\sigma^{\underline\a}(\xi),
$$
where in addition to the component fields already known to the
reader, which obey the relations similar to \p{soro-11} and
\p{soro-12}, there appear auxiliary fields $y^{\underline m}(\xi)$
and $\sigma^{\underline\a}(\xi)$ which are expressed through other
components and/or their derivatives via the superembedding
conditions \p{se2}. We shall not present the explicit form of
these expressions here but just show how the pure spinor
conditions follow from \p{se2}. To this end let us hit the first
expression in \p{se2} by $D$ and the second one by $\bar D$. In
view of the $n=2$ superalgebra commutation relations \p{DbarD} we
have
$$
D\Theta\Gamma^{\underline m}D\Theta=0, \qquad
 \bar D\Theta\Gamma^{\underline m}\bar D\Theta=0,
$$
and hence
\begin{equation}\label{pure}
\lambda\Gamma^{\underline m}\lambda=0, \quad
\bar\lambda\Gamma^{\underline m}\bar\lambda=0.
\end{equation}
By definition complex commuting spinors satisfying eqs. \p{pure}
are called pure spinors.

As in the previous case of the $N=1$, $D=3$ superparticle one can
convince oneself that the following fermionic constraints are of
the first class
\begin{equation}\label{1st}
\lambda^{\underline\a}D_{\underline\a}=
\lambda^{\underline\a}(\pi_{\underline\a}-ip_{\underline
m}(\Gamma^{\underline m}\theta)_{\underline\a})=0\,,
\end{equation}
$$
\bar\lambda^{\underline\a}D_{\underline\a}=
\bar\lambda^{\underline\a}(\pi_{\underline\a}-ip_{\underline
m}(\Gamma^{\underline m}\theta)_{\underline\a})=0.
$$
In the pure spinor approach to the covariant quantization
\cite{Berkovits:2022fth} the spinors $\bar\lambda$ never appear, the
spinors $\lambda$ are ghosts and the constraint
$Q=\lambda^{\underline\a}D_{\underline\a}$ plays the role of the
BRST charge whose cohomology has been shown to reproduce the
correct physical spectrum of quantized superparticles and
superstrings.

We have thus shown that ingredients of the pure spinor
quantization procedure appear in the superembedding formulation.
A relation between the two approaches has been established
in \cite{Matone:2002ft} with the example of the heterotic string. We refer
the reader to this article for further details and pass to
 the superembedding description of an $N=1$, $D=10$ superparticle with $n=8$ worldline
supersymmetry.

\subsection{$N=1$, $D=10$ superparticle with $n=8$ worldline
supersymmetry} 

Let us consider now the double supersymemtric formulation of an $N=1$, $D=10$ superparticle with a worldline superspace having $n=8$ real Grassmann coordinates $\eta^q$, $q=1,...,8$. The worldline fermionic covariant
derivatives are collected in an octuplet of the $SO(8)$ symmetry group
\begin{equation}\label{Dq=n8}
D_q={\partial\over{\partial\eta^q}}+i\eta^q {\partial\over{\partial\xi}}\quad
\end{equation}
and generate the worldline $n=8$ supersymmetry algebra
\begin{equation}\label{DqDp}
 \{D_q,D_p\}=2i\delta_{qp} \partial_\xi \; , \qquad  \partial_\xi:={\partial\over{\partial\xi}}\,.
\end{equation}
These derivatives can be obtained by decomposing
the external differential in worldline superspace
\begin{equation}\label{d=}
d=d\xi \partial_\xi +d\eta^q \partial_q = e \partial_\xi +d\eta^q D_q\; , \qquad e =d\xi -id\eta^q\,\eta^q\; .
\end{equation}
in the
basis of the bosonic and fermionic supervirlbeins
\begin{equation}\label{eA=}
e^A= (e,e^q)=(d\xi -id\eta^q\,\eta^q, \; d\eta^q) \; .
\end{equation}
The bosonic and fermionic coordinate functions are now superfields with nine terms in the decomposition
\begin{eqnarray}\label{X=n8}
X^{\underline m}(\xi,\eta)=x^{\underline
m}(\xi)+i\eta^q\chi_q^{\underline
m}(\xi)+ \ldots + \frac 1 {8!}  \eta^{q_8}\ldots \eta^{q_1} \varepsilon_{q_1\ldots q_8}
y^{\underline m}(\xi)\,,
\\ \label{Th=n8}
\Theta^{\underline\a}(\xi,\eta,\bar\eta)=\theta^{\underline\a}(\xi)+
\eta^q\lambda^{\underline\alpha}_q(\xi)+ \ldots + \frac 1 {8!}  \eta^{q_8}\ldots \eta^{q_1} \varepsilon_{q_1\ldots q_8} \zeta^{\underline\alpha} (\xi) ,
\end{eqnarray}
The superembedding equation has the form
\begin{equation}\label{Eq=0}
E_q^{\underline m}= D_q X^{\underline m} -
iD_q\Theta\Gamma^{\underline m}\Theta =
D_q X^{\underline m} -
iD_q\Theta^{\underline{\alpha}}\Gamma^{\underline m}_{\underline{\alpha}\underline{\beta}}\Theta^{\underline{\beta}}=0\,,
\end{equation}
where the notation $E_q^{\underline m}$ reflects the fact that
the l.h.s. of the superembedding condition is given by the fermionic component of the superworldline pull-back of the $N=1$, $D=10$ Volkov-Akulov one-form
\begin{equation}\label{VA=10I}
E^{\underline m}=d
X^{\underline m}- id\Theta\Gamma^{\underline m}\Theta\; , \qquad
\end{equation}
where now  $\Gamma^{\underline m}_{\underline{\alpha\beta}}$ are $8\times 8 $ symmetric matrices (a D=10 generalization of the relativistic Pauli matrices) obeying \eqref{gammafierz}.

 Using the algebra \eqref{DqDp} we find that selfconsistency conditions for the superembedding equation \eqref{Eq=0},
$D_qE_p^{\underline m}+D_pE_q^{\underline m}=0$, imply
\begin{equation}\label{E0a=DTDT}
D_q\Theta\Gamma^{\underline m}D_p\Theta= \delta_{qp}E_0^{\underline m} \,,
\end{equation}
where
$$
E_0^{\underline m}=\partial_\xi X^{\underline m}  -
i\partial_\xi\Theta^{\underline{\alpha}}\Gamma^{\underline m}_{\underline{\alpha}\underline{\beta}}\Theta^{\underline{\beta}}\,.
$$
Considering this relation for equal $q$ and $p$, for instance for $q=p=1$ we find that, due the to gamma-matrix identities \eqref{gammafierz}, it implies the light-likeness of the ten-vector superfield $E_0^{\underline m}$,
\begin{equation}\label{E0=DT1DT1}
E_0^{\underline m}= D_1\Theta^{\underline{\alpha}}\Gamma^{\underline m}_{\underline{\alpha}\underline{\beta}}D_1\Theta^{\underline{\beta}}
\qquad \Rightarrow \qquad E_{0 \underline m}E_0^{\underline m}=0\,.
\end{equation}

Similarly, the leading component ($\eta^q=0$ part) of the superfield relation \eqref{E0a=DTDT}
\begin{equation}\label{dxa-=ll}
\lambda_q^{\underline{\alpha}}\Gamma^{\underline m}_{\underline{\alpha}\underline{\beta}}\lambda_p^{\underline{\beta}}=
\delta_{qp}(\partial_\xi x^{\underline m}  -
i\partial_\xi\theta\Gamma^{\underline m}\theta) \,
\end{equation}
implies the light-likeness of the superparticle momentum
\begin{equation}\label{pp=0}
p^{\underline m}\propto (\partial_\xi x^{\underline m}  -
i\partial_\xi\theta\Gamma^{\underline m}\theta) \, , \qquad p^{\underline m}p_{\underline m}=0\; .
\end{equation}
 The relation  \eqref{dxa-=ll} thus provides a 10D generalization of  the four-dimensional Cartan-Penrose (twistor) representation of a light-like vector.

Moreover, the relations  \eqref{dxa-=ll} impose strong algebraic constraints on the bosonic spinors
 $\lambda_q^{\underline{\alpha}}$ reducing drastically the number of their independent components. Using the Lorentz-harmonic formalism of \cite{Delduc:1991ir,Galperin:1991gk,Bandos:1991bh} (see \cite{Bandos:2017zap,Bandos:2017eof,Bandos:2019zqp} for a recent discussion) one can show that, taking into account certain gauge symmetries of this model,
 the constrained $\lambda_q^{\underline{\alpha}}$  can be identified with homogeneous coordinates of the space  ${\bb R}\otimes {\bb S}^9$, where ${\bb S}^9$ is a celestial sphere.

It can be shown  \cite{Galperin:1992bw} that the superembedding equation reduces the field content of the coordinate superfields \eqref{X=n8} and \eqref{Th=n8} to the leading components
$x^{\underline{m}}(\xi)$ and $\theta^{\underline{\alpha}}(\xi)$  and to the bosonic spinors $\lambda_q^{\underline{\alpha}}(\xi)$ constrained by \eqref{dxa-=ll}. However, it does not produce equations of motion of the superparticle under consideration. For instance, in view of
\eqref{pp=0}, equation \eqref{dxa-=ll} would be equivalent to the
particle equation of motion $\partial_\xi p^{\underline m}=0$ iff
the bosonic spinors $\lambda_q^{\underline{\alpha}}(\xi)$, or at least their bilinear combination in the l.h.s. of
\eqref{dx-=p} were constant. However, neither this nor the equations of motion for the fermionic coordinate functions $\theta^{\underline{\alpha}}(\xi)$  follow
from the superembedding condition.

The fact that in the case under consideration, as in the cases considered in Sections \ref{D3} and \ref{D10pure}, the superembedding condition does not have dynamical equations among its consequences, implies that there exists a worldline superfield action for the $N=1$, $D=10$ superparticle with manifest $n=8$ worldline supersymmetry \cite{Galperin:1992bw} similar to the actions \eqref{n1d3} and \eqref{n2d10}
\begin{equation}\label{n8d10}
S= \int d\xi d^8\eta P^q_{\underline m}\left(D_q X^{\underline m}-
iD_q\Theta\Gamma^{\underline m}\Theta\right)\,.
\end{equation}
The variation of this action with respect to the Grassmann-odd Lagrange multiplier superfield $P^q_{\underline m}(\xi,\eta^q)$ produces the superembedding condition \eqref{Eq=0} as desired. But a nontrivial problem is to prove that this Lagrange multiplier does not contain superfluous dynamical degrees of freedom. To this end it is important to notice that the action \eqref{n8d10} is invariant under the following local symmetry transformation
\begin{equation}\label{vPqm=DXi}
\delta P^q_{\underline m}= D_p \left(
\Sigma^{qpr \underline \alpha}\Gamma_{\underline m\underline \alpha\underline \beta}D_r\Theta^{\underline \beta}
\right)\; , \qquad
\delta X^{\underline m}=0\; , \qquad
\delta \Theta^{\underline \alpha}=0\,.
\end{equation}
Where the spinor superfield $\Sigma^{qpr \underline \alpha}(\xi,\eta)$ parameter is symmetric and traceless in its $SO(8)$ indices
\begin{equation}\label{Xi=sym}
\Sigma^{qpr \underline \alpha}=
\Sigma^{pqr \underline \alpha}=\Sigma^{(qpr) \underline \alpha}\; , \qquad \Sigma^{qqr \underline \alpha}=0\; . \qquad
\end{equation}
Solving the  equations of motion of the coordinate superfields $X^{\underline m}(\xi,\eta)$ and $\Theta^{\underline m}(\xi,\eta)$
\begin{equation}\label{DqPa=0}
D_qP^q_{\underline m}=0 \; , \qquad  P^q_{\underline m}D_q\Theta\Gamma^{\underline m}=0\,
\end{equation}
and using the above gauge symmetry one can reduce the Lagrange multiplier superfield to   just one term
\begin{equation}\label{Pa=epa}
P^q_{\underline m}=\frac 1 {7!}\varepsilon^{qp_1\ldots p_7}
\eta^{p_1}\ldots \eta^{p_7} p_{\underline m}
\, ,
\end{equation}
where $p_{\underline m}$ is a constant vector
\begin{equation}\label{dpa=0}
\frac d{d\xi} p_{\underline m}=0
\, .
\end{equation}
It is expressed through the bilinear combinations of the bosonic spinors $\lambda_q^{\underline{\alpha}}$ and hence is light-like
\begin{equation}\label{pa=ll}
 p_{\underline m}\propto \lambda_q\Gamma_{\underline m}\lambda_q
\qquad \Rightarrow \qquad  p_{\underline m} p^{\underline m}=0\; .
\end{equation}
Furthermore, taking into account the consequence \eqref{dxa-=ll} of the superembedding condition \eqref{Eq=0}, we conclude that \eqref{pa=ll} implies
\begin{equation}\label{dx-=p}
\partial_\xi x^{\underline m}  -
i\partial_\xi\theta\Gamma^{\underline m}\theta =ep^{\underline m}
\end{equation}
with some function $e(\xi)$ which can be gauge fixed to a constant using the (super)conformal symmetry of the action
\eqref{n8d10}.   This completes the proof of the equivalence of the doubly supersymmetric model described by the action \eqref{n8d10} and the Green-Schwarz-like  formulation of the $N=1$, $D=10$ massless superparticle mechanics, with the $\kappa$-symmetry being related to the $n=8$ local worldline supersymmetry in a way similar to that discussed in Section \ref{ksusy}.

Note that in the description of type IIA and type IIB $D=10$ superparticles, superstrings and D-p-branes in worldvolume superspace with $n=16$ supersymmetry (which is the number of independent $\kappa$-symmetries in these cases) the superembedding condition puts the theories on the mass shell, i.e. it contains the dynamical equations among its consequences. As a result one cannot use an $n=16$ superfield generalization of the action
\eqref{n8d10} to describe type IIA and type IIB super-p-branes. In these cases the superfield equations can be obtained from the generalized action principle \cite{Bandos:1995dw,
Bandos:1997rq,Howe:1998tsa} which is a super-p-brane counterpart of the so-called rheonomic (group manifold) approach to supergravity \cite{Neeman:1978njh,Neeman:1978zvv,D'Auria:1982nx,Castellani:1991et}.

\section{Superembedding description of superstrings}

\subsection{$N=1$, $D=10$ (heterotic) superstring}

To replace all the $\kappa$-symmetries of the  $N=1$, $D=10$ superstring with manifest worldsheet supersymmetry, we should introduce an $n=(8,0)$ worldlsheet superspace  parametrized by coordinates
\begin{equation}\label{zM=8+0}
z^{M}= (\xi^m,\eta^q)=(\xi^{+},\xi^{-}, \eta^{q})\; , \qquad q=1,\ldots,8\,,
\end{equation}
where $\xi^{\pm}$ are light-cone coordinates of the worldsheet which are transformed under the $2d$ Lorentz group $SO(1,1)$ with the Lorentz weights $\pm 1$, respectively, while $\eta^q$ are Grassmann-odd $2d$ Majorana-Weyl spinors which are transformed under $SO(1,1)$ with the Lorentz weight $+\frac 12$, i.e. they are "left-handed".

We consider the following action on the superworldsheet coordinates of a superconformal symmetry parametrized by the superfield  $\Lambda^{+}(\xi^{+}, \eta^{q})$ depending on the "left-handed" bosonic and fermionic coordinates only:
\begin{equation}\label{2dsusy}
\delta \xi^{+}= \Lambda^+ - \frac 1 2 \eta^{q}D_{q}\Lambda^{+}\; , \qquad   \delta \eta^{q}= - \frac i 2 D_{q}\Lambda^{+}\; , \qquad \delta \xi^{-}=0\; .\qquad
\end{equation}
Note that  the "right-moving" coordinate $\xi^-$ is inert under these supersymmetry transformations.

In \eqref{2dsusy}, $D_{q}$ are the fermionic covariant derivatives (similar to
\eqref{Dq=n8})
$$
D_{q}= \partial_{q}+i  \eta^{q} \partial_{+}\; , 
$$
that obey the $n=(8,0)$ supersymmetry algebra
\begin{equation}\label{D+qD+p}
 \{D_{q},D_{p}\}=2i\delta_{qp} \partial_{+} \;.
\end{equation}
A key property of the superconformal symmetry is the homogeneous transformation of the covariant derivatives
$$ \delta D_{q} =  \frac i 2 (D_{q}D_p \Lambda^+) \, D_{p} \;.
$$
The other object which transforms homogeneously under this superconformal symmetry is the
 Volkov-Akulov vielbein of the flat $d=2$, $n=(8,0)$ superspace
\begin{equation}\label{VA=2d8n}
e^{+}=d\xi^{+} -id\eta^{q}\, \eta^{q}\; ,
\end{equation}
$$ \delta e^{+}= e^+ \partial_+ \Lambda^{+} \; . \qquad  $$
The superfield parametrizing these superconformal transformations appear as formal contraction of this VA 1-form with the variation symbol,
$$
i_\delta e^{+}:=\delta \xi^{+} -i\delta \eta^{q}\, \eta^{q}=\Lambda^+\; .
$$

The coordinate functions are now  worldsheet superfields
\begin{eqnarray}\label{X=n+8}
X^{\underline m}(\xi^m,\eta^p)&=&x^{\underline
m}(\xi^m)+i\eta^q\chi_{q}^{\underline
m}(\xi^m)+ \ldots + \frac 1 {8!}\eta^{q_8}\ldots \eta^{q_1} \varepsilon_{q_1\ldots q_8}\,
y^{\underline m}(\xi^m)\,,
\\ \label{Th=n+8}
\Theta^{\underline\a}(\xi^m,\eta^p)&=&\theta^{\underline\a}(\xi^m)+
\eta^{q}\lambda^{\underline\alpha}_q(\xi^m)+ \ldots + \frac 1 {8!}\eta^{q_8}\ldots \eta^{q_1} \varepsilon_{q_1\ldots q_8}\,\zeta^{\underline\alpha} (\xi^m).
\end{eqnarray}

The superembedding condition is again a particular case of
\eqref{flatse}
\begin{equation}\label{E+q=0}
E_{q}^{\underline m}= D_{q} X^{\underline m} -
iD_{q}\Theta\Gamma^{\underline m}\Theta =0\,.
\end{equation}
Which implies that the pullback of the one-form $E^{\underline m}= d X^{\underline m} -
id\Theta\Gamma^{\underline m}\Theta$ is
\begin{equation}\label{Ea=e++E++}
E^{\underline m}= e^{+} \, E_{+}^{\underline m} + e^{-} \, E_{-}^{\underline m}\; , \qquad
\end{equation}
where
\begin{equation}
\label{E++a:=} E_{+}^{\underline m}=\partial_{+} X^{\underline m}  -
i\partial_{+}\Theta^{\underline{\alpha}}\Gamma^{\underline m}_{\underline{\alpha}\underline{\beta}}\Theta^{\underline{\beta}}\; , \qquad
E_{-}^{\underline m}=\partial_{-} X^{\underline m}-i\partial_{-}\Theta^{\underline{\alpha}}\Gamma^{\underline m}_{\underline{\alpha}\underline{\beta}}\Theta^{\underline{\beta}} \; .  \qquad
\end{equation}

As in the superparticle case, studying the selfconsistency conditions for the superembedding equation \eqref{E+q=0}
 using the algebra \eqref{D+qD+p} we find that
\begin{equation}\label{E++a=DTDT}
D_{q}\Theta\Gamma^{\underline m}D_{p}\Theta= \delta_{qp}E_{+}^{\underline m} \,
\end{equation}
which implies
\begin{equation}\label{E++2=0}
 E_{+\underline m}E_{+}^{\underline m}=0\,.
\end{equation}
Similarly, the leading component  of  \eqref{E++a=DTDT},
\begin{equation}\label{d++xa-=ll}
\lambda_q^{\underline{\alpha}}\Gamma^{\underline m}_{\underline{\alpha}\underline{\beta}}\lambda_p^{\underline{\beta}}=
\delta_{qp}(\partial_{+} x^{\underline m}  -
i\partial_{+}\theta\Gamma^{\underline m}\theta) \,
\end{equation}
implies the left-handed Visrasoro constraint
\begin{equation}\label{d++x2=0}
 (\partial_{+} x^{\underline m}  -
i\partial_{+}\theta\Gamma^{\underline m}\theta)^2=0 \, ,
\end{equation}
the counterpart of the light-likeness of the momentum of the massless superparticle \eqref{pp=0}.

As in the case of the $N=1$, $n=8$ superparticle, the relations  \eqref{d++xa-=ll} impose strong algebraic constraints on the bosonic spinors  $\lambda_q^{\underline{\alpha}}$ making possible to  identify them, up to a scalar multiplier, with Lorentz-spinor harmonic variables.
Furthermore, similar to the superparticle case, the superembedding condition reduces the dynamical field content of the superfields \eqref{X=n+8} and \eqref{Th=n+8} to their leading components,
$x^{\underline{m}}(\xi^{\pm})$ and $\theta^{\underline{\alpha}}(\xi^{\pm})$, and  the bosonic spinors $\lambda_q^{\underline{\alpha}}(\xi^{\pm})$ constrained by \eqref{d++xa-=ll} {\cite{Delduc:1992fk}}. The dynamical equations of motion of these fields are not part of the superembedding condition, so one can construct an $n=(8,0)$ worldsheet superspace action which will produce the superembedding condition and the dynamical equations of motion of the $N=1$, $D=10$ superstring.

Notice however, that the right-handed Virasoro constraint of the string theory
\begin{equation}\label{d--x2=0}
(\partial_{-} x^{\underline m}-i\partial_-\theta\Gamma^{\underline m}\theta)^2=0
\end{equation}
does not follow from the superembedding condition \eqref{E+q=0}.
To obtain the second Virasoro constraint from the superstring action one should consider
a slightly curved geometry on the superworldsheet $\mathcal M_{sw}$  which is described by the following worldsheet supervielbein
\begin{eqnarray}
&& e^A= (e^+,\,e^-,\,e^{q}),\qquad \nonumber
\\ \label{e--=}
&& e^{+} =d\xi^+- i d\eta^{q}\,\eta^{q} \ \;,   \qquad e^{q}=d\eta^{q}\; , \qquad e^{-}=d\xi^-+\frac i 8 e^+ D_{q} e^-_q +  e^{q}\, e^-_q  \;.   \qquad
\end{eqnarray}
It contains the same flat supervilebeins in the left-handed sector and a  more complicated right-handed einbein $e^-$ whose components are expressed in terms of the  superfield $e^-_q(\xi^m,\eta^p)$ which obeys the constraint
\begin{equation}\label{Dmu=1}
D_{(p}e_{q)}{}^{\!\!\! -} = \frac 1 8\,  \delta_{qp} D_{r}e_{r}{}^{\!\!\! -}\; .
\end{equation}
The worldovolume component $D_{r}e^-_{r}\vert_{\eta^q=0}=e^{--}(\xi^m)$ is an auxiliary field whose equation of motion obtained from the action \eqref{n8d10=str} considered below will produce the left-handed Virasoro constraint \eqref{d--x2=0}.

 The corresponding covariant derivatives are
$$
\nabla_{q}= D_{q}- e^-_q\partial_- \; , \qquad \nabla_{+}= \partial_{+}+\frac i 8 D_{q} e^-_q\partial_-\; , \qquad \nabla_{-}= \partial_{-} \;.  \qquad
$$
They form a `quasi flat' superalgebra {\textcolor{blue}{with}}
$$
{} \{\nabla_{q},\nabla_{p}\}= 2i\delta_{qp} \nabla_{+}, \qquad [\nabla_+,\nabla_q]=0\,.
$$
For this choice of the worldsheet supergeometry the superembedding condition takes the form
\begin{equation}\label{senew}
E_{q}^{\underline m}= \nabla_{q} X^{\underline m} -
i\nabla_{q}\Theta\Gamma^{\underline m}\Theta =0\,.
\end{equation}

With this superworldsheet geometry at hand the authors of
\cite{Delduc:1992fk} proposed the following  $n=(8,0)$ $2d$ superfield  action for the description of $N=1$, $D=10$ superstring
\begin{equation}\label{n8d10=str}
S= \int d^2\xi d^8\eta P^{q}_{\underline m}\left(\nabla_{q} X^{\underline m}-
i\nabla_{q}\Theta\Gamma^{\underline m}\Theta\right) + \int d^2\xi d^8\eta {\cal P}^{MN}\left({\cal L}_{MN} -\partial_MY_N\right)\,.
\end{equation}
The first term in this action has the structure similar to the superparticle action \eqref{n8d10}, the integral of the superembedding condition multiplied by Lagrange multiplier superfield. This term is invariant under local transformations of the Lagrange multiplier similar to those in \eqref{vPqm=DXi}.
The second term is new, it is required to generate the string tension and reproduce the Wess-Zumino term of the Green-Schwarz formulation of the superstring. Here ${\cal P}^{MN}= -(-)^{MN}{\cal P}^{NM}$ is the graded-(anti)symmetric Lagrange multiplier\footnote{$(-)^{MN}$ is a shortcut notation for $(-1)^{\epsilon(M)\epsilon (N)}$ where $ \epsilon(M)=\epsilon(z^M)$ is the Grassmann parity equal to 0 for the bosons and to 1 for the fermions. 
},
 $Y_M(z)$ is an auxiliary superfield and $\mathcal L_{MN}$ are components of the worldsheet superform ${\cal L}_2= \frac 1 2 dz^N \wedge dz^M {\cal L}_{MN}$:
\begin{equation}\label{cL2=n8}
    {\cal L}_2 = B_2 +e^+\wedge e^- E_+^{\underline{m}} E_{-\underline{m}} \; , \qquad
\end{equation}
where
\begin{equation}\label{B2=}
    B_2= -i E^{\underline{m}}\wedge d\Theta \Gamma_{\underline{m}}\Theta
\end{equation}
is the pullback of the so-called NS-NS 2-form of flat $N=1$, $D=10$ superspace, and
$E_+^{\underline{m}}$ and $ E_-^{\underline{m}}$ are components of the decomposition of the pullback of the target superspace vector supervielbein \eqref{VA=10I} in the covariant basis \eqref{e--=}
\begin{equation}
E^{\underline{m}}=  e^+E_+^{\underline{m}}+ e^-  E_-^{\underline{m}} +e^{q}E_{q}^{\underline{m}}\;.  \qquad
\end{equation}

 An important property of the 2-form \eqref{cL2=n8} is that it is closed if the superembedding condition \eqref{senew} is satisfied
 \begin{equation}\label{closer}
d{\cal L}_2\vert_{E_q^{\underline{m}}=0}=0\;  \qquad \rightarrow
\partial_{[M}{\cal L}_{NK)}\vert_{E_q^{\underline{m}}=0}=0\; , \qquad
\end{equation}
(where $[\ldots)$ denotes the Grassmann-graded (anti)symmetrization of the indices).

This implies that the action is invariant under a gauge symmetry
with the superfield parameter
$$ \Sigma^{MNK}(z)= -(-)^{MN}\Sigma^{NMK}(z) =-(-)^{NK}\Sigma^{MKN}(z) =  \Sigma^{[MNK)}(z)$$
which acts on the Lagrange multiplier ${\cal P}^{MN}$ as follows
 \begin{equation}\label{vcPMN=}
\delta {\cal P}^{MN} = (-)^K\partial_K \Sigma^{MNK}(z)\; .  \qquad
\end{equation}
Due to the closer of $\mathcal L_{MN}$ modulo the superembedding condition \eqref{closer}, the variation of the action with respect to \eqref{vcPMN=} will be proportional to the superembedding condition and can be therefore compensated by an appropriate transformation of the Lagrnage multiplier $P^{+q}_{\underline{m}}$.

The symmetry \eqref{vcPMN=} allows to gauge fix the general solution of the equations of motion for the auxiliary superfield
$Y_M(z)$,
$$
(-)^M\partial_M{\cal P}^{MN}=0
$$
to the simple expression
\begin{equation}\label{PMN=T}
{\cal P}^{MN}=\frac 1 {8!}\eta^{q_8}\ldots \eta^{q_1} \varepsilon_{q_1\ldots q_8}\,e^M_+\,e^M_- T
\end{equation}
with a constant $T$ which has the meaning of the superstring tension. Thus the action \eqref{n8d10=str} realizes a dynamical tension generation mechanism for supestrings put forward in \cite{Townsend:1992fa,Bergshoeff:1992gq}
(see \cite{Bandos:1993pa} for more details). It can be shown that the action \eqref{n8d10=str} is classical equivalent to the Green-Scwharz action for an $N=1$, $D=10$ superstring. For a detailed demonstration of this fact we refer the reader to  \cite{Delduc:1992fk,Sorokin:1999jx}, while here we only note that if we integrate the action \eqref{n8d10=str} over the fermionic variables $\eta^q$, impose a conformal gauge on the worldsheet supervielbeins, eliminate all the auxiliary fields by using the local symmetry transformations \eqref{vPqm=DXi} and \eqref{vcPMN=}, and by solving the superembedding condition, then the action \eqref{n8d10=str} reduces to the following one
\begin{eqnarray}\label{GS}
T\int {\cal L}_2\vert_{\eta^{q} =0} &= &T\int  d^2\xi  (\partial_+ x^{\underline{m}}-i \partial_+\theta\Gamma^{\underline{m}}\theta) (\partial_- x_{\underline{m}}-i\partial_-\theta\Gamma^{\underline m}\theta ) \nonumber\\
&&-i\, T\int  (d x^{\underline{m}}-i d\theta\Gamma_{\underline{m}}\theta)\wedge d\theta \Gamma_{\underline{m}}\theta\;,
\end{eqnarray}
This is the action, in the conformal gauge, for the $N=1$, $D=10$ Green-Schwarz superstring in flat target superspace.

To describe within the superembedding approach a fully fledged heterotic string  \cite{Gross:1984dd} carrying on its worldsheet 32 chiral fermions or 16 chiral bosons generating  an $SO(32)$ or $E_8\times E_8$ gauge group, one should add to the action \eqref{n8d10=str} terms which describe the dynamics of these heterotic degrees of freedom. The construction of such terms with the use of $n=8$ worldsheet superfields turned out to be not a straightforward enterprise. For different ways of realizing the chiral fermion  construction we refer the reader to \cite{Sorokin:1993pm,Howe:1994he,Ivanov:1994fc}.

\subsection{$N=2$, $D=10$ superstrings with $n=(8,8)$ worldsheet supersymmetry. Equations of motion from the superembedding condition}

Let us now consider examples of superstrings whose superembedding description produces all the equations of motion, i.e. is entirely on-shell formulation. This is the case of type IIA and type IIB $D=10$ superstrings.

The $N=2$, $D=10$ target superspace of the type IIA superstring has 32 fermionic directions parametrized by
two Majorana-Weyl spinor coordinates of opposite chiralities
\begin{equation}
    IIA:\qquad Z^{\underline{M}}= ( X^{\underline{m}}, \Theta ^{\underline{\alpha}1}, \Theta ^2_{\underline{\alpha}})\; , \qquad \underline{\alpha}=1,...,16
\end{equation}
and in the type IIB superstring case, 32 fermionic directions of the target superspace are parametrized by two Majorana-Weyl spinor coordinates of the same chirality
\begin{equation}\label{IIB}
    IIB:\qquad Z^{\underline{M}}= ( X^{\underline{m}}, \Theta ^{\underline{\alpha}1},  \Theta ^{\underline{\alpha}2})\; , \qquad \underline{\alpha}=1,...,16\; .
\end{equation}
Since the superembedding descriptions of the two cases are quite similar, here we will only discuss the type IIB case, because, as we will show below, in this case the superembedding approach  provides  us with a universal description of the fundamental type IIB superstring and a Dirichlet string also called the super-D1-brane \cite{Bandos:1995zw,Bandos:2000hw}.

To replace all the 16 kappa-symmetries of the GS formulation of the type IIB superstring with the manifest worldsheet supersymmetry, we shall consider an embedding into the type IIB target superspace of an $n=(8,8)$ worldsheet superspace $\mathcal M_{sw}$ parametrized by coordinates
$z^M= (\xi^{+},\xi^{-},\eta^{q}, \eta^{\dot q})$. The eight fermionic coordinates $\eta^q$ ($q=1,\ldots,8$) are `left-handed', they are Grassmann-
odd $2d$ Majorana-Weyl spinors which are transformed under $SO(1, 1)$ with the Lorentz
weight $+\frac 12$. The eight fermionic coordinates $\eta^{\dot q}$ are `right-handed', they are Grassmann-
odd $2d$ Majorana-Weyl spinors which are transformed under $SO(1, 1)$ with the Lorentz
weight $-\frac 12$. $\eta^q$ and $\eta^{\dot q}$ transform under different  8-dimensional (s- and c-spinor) representations of $SO(8)$ related to each other by triality.

For simplicity we will consider the case of the superembedding of the flat superworldsheet. This will result in the on-shell description of the type IIB superstring in the conformal gauge.
To get from the superembedding the type IIB superstring equations of motion which are not subject to the conformal gauge, one should deal with a geometry of a curved $n=(8,8)$  superworldsheet (see \cite{Bandos:1995zw} and \cite{Bandos:1995dw,Bandos:2000hw,Bandos:2008ba} for more details in a bit different setups).

The supersymmetry invariant supervielbeins of the flat $n=(8,8)$  superworldsheet are
\begin{eqnarray}\label{88sv}
    e^{+}=d\xi^{+}-id\eta^{q}\eta^{q}, \qquad e^{q}=d\eta^{q}\; , \qquad \nonumber\\
    e^{-}=d\xi^{-}-id\eta^{\dot q}\eta^{\dot q}, \qquad e^{\dot q}=d\eta^{\dot q}\; . \qquad
\end{eqnarray}
The corresponding fermionic covariant derivatives are
\begin{equation}\label{D+q=}
D_{q}= \partial_{q}+i  \eta^{q} \partial_{+}\; , \qquad D_{\dot{q}}= \partial_{\dot{q}}+i  \eta^{\dot{q}}\partial_{-}\; . \qquad
\end{equation}
They obey the rigid $n=(8,8)$ supersymmetry algebra
\begin{equation}\label{D+qD-p}
 \{D_{q},D_{p}\}=2i\delta_{qp} \partial_+ \; , \qquad  \{D_{q},D_{\dot p}\}=0 \; , \qquad  \{D_{\dot{q}},D_{\dot{p}}\}=2i\delta_{\dot{q}\dot{p}} \partial_{-} \; .  \qquad
\end{equation}
We require that the $n=(8,8)$ superworldsheet is embedded into a flat type IIB $D=10$ superspace \eqref{IIB} in such a way that the pull-back of the Volkov-Akulov one-form
\begin{equation}\label{VA=IIB}
E^{\underline a}=dZ^{\underline{M}}E_{\underline{M}}^{\underline{a}}=d
X^{\underline a}- id\Theta^1\Gamma^{\underline a}\Theta^1- id\Theta^2\Gamma^{\underline a}\Theta^2\;  \qquad
\end{equation}
on the tangent space basis \eqref{88sv} of the superworldsheet is non-zero only along the bosonic directions
\begin{eqnarray}\label{Ea=IIBsEmb}
E^{\underline m}&=&e^{+} E_{+}^{\underline m} + e^{-} E_{-}^{\underline m}\\
&=& e^{+}(\partial_+
X^{\underline m}- \partial_+\Theta^1\Gamma^{\underline m}\Theta^1- i\partial_+\Theta^2\Gamma^{\underline m}\Theta^2) +e^{-}(\partial_-
X^{\underline m}- \partial_-\Theta^1\Gamma^{\underline m}\Theta^1- i\partial_-\Theta^2\Gamma^{\underline m}\Theta^2)\,.\nonumber
\end{eqnarray}
This implies the conventional form of the superembedding conditions
\begin{eqnarray}\label{E+qIIB=0}
E_{q}^{\underline m}=D_{q}
X^{\underline m}- D_{q}\Theta^1\Gamma^{\underline m}\Theta^1- iD_{q}\Theta^2\Gamma^{\underline m}\Theta^2=0  \; , \qquad\nonumber \\
E_{\dot q}^{\underline m}=D_{\dot q}
X^{\underline m}- D_{\dot q}\Theta^1\Gamma^{\underline m}\Theta^1- iD_{\dot q}\Theta^2\Gamma^{\underline m}\Theta^2 =0 \; . \qquad
\end{eqnarray}
The self-consistency conditions which follow from \eqref{E+qIIB=0} upon applying to the latter the covariant derivatives $D_p$ and $D_{\dot p}$ and symmetrizing the $SO(8)$ indices  are
\begin{eqnarray}\label{D+E+=0}
    D_{q} E_{p}^{\underline m} + D_{p} E_{q}^{\underline m}=0\; , \qquad   D_{\dot q} E_{\dot p}^{\underline m} + D_{\dot p} E_{\dot q}^{\underline m}=0\; , \qquad
\end{eqnarray}
and
\begin{eqnarray}\label{D+E-+=0}
    D_{q}  E_{\dot p}^{\underline m} + D_{\dot p} E_{q}^{\underline m}=0\; . \qquad
\end{eqnarray}
In view of the anti-commutation relations \eqref{D+qD-p} these conditions take the following form
\begin{eqnarray}\label{E++a=DTDT+}
\delta_{qp}E_{+}^{\underline m} = D_{q}\Theta^1\Gamma^{\underline m}D_{p}\Theta^1 +D_{q}\Theta^2\Gamma^{\underline m}D_{p}\Theta^2  \, , \qquad \\
\label{E--a=DTDT-}
\delta_{\dot q\dot p}E_{-}^{\underline m} = D_{\dot q}\Theta^1\Gamma^{\underline m}D_{\dot p}\Theta^1 +D_{\dot q}\Theta^2\Gamma^{\underline m}D_{\dot p}\Theta^2  \, , \qquad
\end{eqnarray}
and
\begin{eqnarray}\label{D-TD+T+=0}
D_{q}\Theta^1\Gamma^{\underline m}D_{\dot p}\Theta^1 +D_{q}\Theta^2\Gamma^{\underline m}D_{\dot p}\Theta^2 =0  \, . \qquad
\end{eqnarray}

\subsubsection{Fundamental string solution of the superembedding conditions}
The simplest particular solution of equation \eqref{D-TD+T+=0} is
\begin{eqnarray}\label{D-T1=0}
D_{\dot p}\Theta^{1\underline{\alpha}}=0 \; , \qquad  D_{p}\Theta^{2\underline{\alpha}} =0  \,  \qquad
\end{eqnarray}
which, in view of the superalgebra \eqref{D+qD-p} implies
\begin{eqnarray}\label{d--T1=0}
\partial_{-}\Theta^{1\underline{\alpha}}=0 \; , \qquad  \partial_{+}\Theta^{2\underline{\alpha}} =0 \,. \,  \qquad
\end{eqnarray}
So the leading components of the fermionic superfields $\Theta^{1,2}|_{\eta=0}=\theta^{1,2}(\xi^m)$
obey the same equations of motion as the fermionic coordinate functions of the type IIB Green-Schwarz superstring formulation in the conformal gauge.

Then  \eqref{E++a=DTDT+} and  \eqref{E--a=DTDT-} reduce to
\begin{eqnarray}\label{E++a=D+TD+T}
\delta_{qp}E_{+}^{\underline m} = D_{q}\Theta^1\Gamma^{\underline m}D_{p}\Theta^1  \, , \qquad \\
\label{E--a=D-TD-T}
\delta_{\dot q\dot p}E_{-}^{\underline m} = D_{\dot q}\Theta^2\Gamma^{\underline m}D_{\dot p}\Theta^2  \, , \qquad
\end{eqnarray}
which (as a consequence of the Fierz identities \eqref{gammafierz}) imply the left- and right-handed Virasoro constraints
\begin{equation}
E_{+}^{\underline m}E_{+ \underline m}=0\; , \qquad E_{-}^{\underline m}E_{-\underline m}=0\; , \qquad
\end{equation}
where, in view of \eqref{d--T1=0}, $E_{+}^{\underline m}$ and $E_{-}^{\underline m}$ reduce to
\begin{eqnarray} \label{E++a:=IIBon}
E_{+}^{\underline m}=\partial_{+} X^{\underline m}  -
i\partial_{+}\Theta^{\underline{\alpha}1}\Gamma^{\underline m}_{\underline{\alpha}\underline{\beta}}\Theta^{\underline{\beta}1}\; , \qquad
E_{-}^{\underline m}=\partial_{-} X^{\underline m} -
i\partial_{-}\Theta^{\underline{\alpha}2}\Gamma^{\underline m}_{\underline{\alpha}\underline{\beta}}\Theta^{\underline{\beta}2}\; .  \qquad
\end{eqnarray}
Acting on eq. \eqref{E++a=D+TD+T} with $\partial_-$ and on \eqref{E--a=D-TD-T} with $\partial_+$, and using \eqref{d--T1=0}we get the equation of motion of the bosonic vector  superfield $X^{\underline m}$
\begin{eqnarray}\label{ddX=0}
\partial_{-}\partial_{+}X^{\underline m} (\xi,\eta^q,\eta^{\dot q})= 0 \qquad \Rightarrow \qquad \partial_{-}\partial_{+}x^{\underline m}(\xi) = 0\,.  \qquad
\end{eqnarray}
Its leading component coincides with the bosonic field equation of the type IIB superstring in the conformal gauge.

\subsubsection{D1-brane solution of the superembedding condition}

As was observed in \cite{Bandos:1995zw},
eq. \eqref{D-T1=0} does not describe the general solution of eq. \eqref{D-TD+T+=0}. The general solution is an $O(2)$ rotated version of \eqref{D-T1=0}
\begin{eqnarray}\label{D-T1+D-T2=0}
\cos \phi\, D_{\dot p}\Theta^{1\underline{\alpha}}- \sin \phi\, D_{\dot p}\Theta^{2\underline{\alpha}}
=0 \; , \qquad  \sin \phi\, D_{p}\Theta^{1\underline{\alpha}}+ \cos \phi\, D_{p}\Theta^{2\underline{\alpha}} =0  , \,  \qquad
\end{eqnarray}
where, {\it a priori}, $\phi$ is a worldsheet superfield. The study of the self-consistency conditions of this solution carried out in \cite{Bandos:1995zw} showed that $\phi$ is actually a constant angle. In \cite{Bandos:2000hw} the angle $\phi$ was related to the on-shell constant value of the field strength $F_{mn}=\partial_mA_n-\partial_nA_m$ of a $2d$ Born-Infeld gauge field living on the wordlsheet of a D1-brane, namely
$$
\tan \phi =\sqrt{\frac {1-F^{(0)}}{1+F^{(0)}}}\; , \qquad F_{0}=\frac 12\epsilon^{mn}{F_{mn}}\,.
$$
Therefore, the solution \eqref{D-T1+D-T2=0} of the superembedding conditions \eqref{E+qIIB=0} provides us with the description of the on-shell dynamics of a super-D1-brane of type IIB string theory (see \cite{Bandos:2000hw} for more details).

To summarize, the geometrical condition on the embedding of an $n=(8,8)$, $d=2$ supersurface into a type IIA or IIB $D=10$ superspace provides us with the full set of constraints and dynamical equations of motion for the on-shell description of type II superstrings and D1-branes. This is also the case for the Dp-branes in type II $D=10$ supergravities and the M2- and M5-branes in $D=11$ supergravity. It is in this way that the equations of motion of the Dp-branes and the M5-brane were first obtained in  \cite{Howe:1996mx} and \cite{Howe:1996yn}  earlier than the actions for these objects were  constructed in \cite{Cederwall:1996ri,Aganagic:1996pe,Bergshoeff:1996tu,Bandos:1997ui,Aganagic:1997zq,Bandos:1997rq}.

Let us note that in the cases in which the superembedding condition puts the theory on the mass-shell, it is not possible to construct worldwolume superfield actions similar to those which we presented for $N=1$ superparticles and superstrings. Instead, in these cases one can use a generalized action principle \cite{Bandos:1995dw,
Bandos:1997rq,Howe:1998tsa} which is a super-p-brane counterpart of the so-called rheonomic (group manifold) approach to supergravity \cite{Neeman:1978njh,Neeman:1978zvv,D'Auria:1982nx,Castellani:1991et}.

\section{Superembedding description of M2 and M5 branes}\label{25}
Let us now consider how the universal superembedding condition \eqref{soro-22} produces the full set of the equations of motion of the M-theory membrane \cite{Bandos:1995zw,Bandos:1995dw}  and the five-brane \cite{Howe:1996yn,Howe:1997fb}.

Now the target superspace is that of $D=11$ supergravity parametrized by the supercoordinates $Z^{\underline M}=(X^{\underline m}, \Theta^{\underline \mu}$), where  $X^{\underline m}$ $(\underline m=0,1,\ldots, 10)$ are eleven bosonic space-time coordinates and $\Theta^{\underline \mu}$ $(\underline\mu=1,\ldots,32)$ are 32 fermionic (Majorana spinor) coordinates. The geometry of this superspace is described by bosonic and fermionic supervielbeins
\begin{equation}\label{11dsv}
E^{\underline a}(Z)=dZ^{\underline M}E_{\underline M}^{\underline a}{\textcolor{blue}{(Z)}}, \qquad E^{\underline \alpha}(Z)=dZ^{\underline M}E_{\underline M}^{\underline \alpha}{\textcolor{blue}{(Z)}},
\end{equation}
and a spin connection one-form $\Omega^{\underline a\underline b}(Z)=dZ^{\underline M}\Omega_{\underline M}^{\underline a\underline b}{\textcolor{blue}{(Z)}}$, which is antisymmetric in the indices $\underline a$ and $\underline b$ and takes values in the vector representation of the local tangent-space (structure) Lorentz group $SO(1,10)$.

The geometry of the $D=11$ superspace is assumed to satisfy the eleven-dimensional supergravity constraints \cite{Cremmer:1980ru,Brink:1980az}, one of the essential ones being the constraint on the bosonic torsion 2-form of the target superspace  (in what follows we will not explicitly right  the wedge product $\wedge$ of differential forms) \footnote{In our conventions the external differential acts on the wedge-product of the differential forms from the right, e.g. in the case of the wedge product of an m-form with an n-form $d(A_m\wedge B_n)=A_m\wedge dB_n+(-1)^n\,dA_m\wedge B_n$.
Let us also recall that for the wedge product of two superforms   $A_m\wedge B_n= (-1)^{mn}(-1)^{[A][B]}B_n\wedge A_m$, where $[A]$ and $[B]$ stand, respectively,  for the Grassmann parity of the forms $A_m$ and $B_n$.}
\begin{equation}\label{Ta}
T^{\underline a}=dE^{\underline a}-E^{\underline b}\,\Omega_{\underline b}{}^{\underline  a}=-iE^{\underline \alpha}\Gamma^{\underline a}_{\underline \alpha\underline \beta}E^{\underline \beta},
\end{equation}
where $\Gamma^{\underline a}_{\underline \alpha\underline \beta}$ are  the $D=11$ Dirac matrices in the Majorana representation (see Appendices A and B).

As is well known, one of the reasons to impose the constraint \eqref{Ta} is that in flat limit it describes the geometry of flat superspace with torsion. In flat superspace the spin connection is flat, i.e. its curvature $R=d\Omega+\Omega\wedge \Omega$ is zero, so the connection is a pure gauge with respect to local transformations of the structure group $SO(1,10)$. Namely,
\begin{equation}\label{Omegaflat}
\Omega_{\underline a}{}^{\underline b}\vert_{ flat}=u_{\;\;\underline a}^{-1\underline c}du_{\underline c}{}^{\underline b}\,
\end{equation}
is the $SO(1,10)$ group Cartan form, where $u_{\underline a}{}^{\underline b}(Z)$ are $SO(1,10)$ matrices (whose  counterparts on $\mathcal M_{sw}$, called Lorentz harmonics, will play an important role in our superembedding description)
\begin{equation}\label{u}
u_{\underline a}{}^{\underline c}u_{\underline a}{}^{\underline d}\eta_{\underline {cd}}=\eta_{\underline{ab}}\,,\qquad
u_{\underline a}{}^{\underline c}u_{\underline a}{}^{\underline d}\eta^{\underline{ab}}=\eta^{\underline{cd}}\,,\qquad u^{-1}=u^T,
\end{equation}
and $\eta_{\underline{ab}}$ is the $D=11$ Minkowski metric whose signature is chosen to be `almost minus'.

Thus, by an appropriate local Lorentz transformation which acts on the vector and spinor components of the supervielbein \eqref{11dsv}, in flat superspace one can set $\Omega_{\underline a}{}^{\underline b}=0$. In this basis the vector supervielbein of the flat superspace is the Volkov-Akulov one-form \eqref{VA1f},
\begin{equation}\label{Ea=flat11D}
E^{\underline a}=dX^{\underline a}-id\Theta^{\underline \alpha}\Gamma^{\underline a}_{\underline \alpha\underline\beta}\Theta^{\underline \beta},
\end{equation}
  the spinor supervielbein is the exact 1--form $E^{\underline \alpha}=d\Theta^{\underline \alpha}$ and the torsion \eqref{Ta} reduces to
\begin{equation}\label{Taflat}
T^{\underline a}_{flat}=-id\Theta^{\underline \alpha}\Gamma^{\underline a}_{\underline \alpha\underline\beta}d\Theta^{\underline \beta}.
\end{equation}
Though the superembedding approach has been developed in full generality for the description of branes propagating in curved target superspaces of supergravity theories (see e.g. \cite{Sorokin:1999jx,Bandos:2009xy} for a detailed review), for simplicity in what follows we will restrict our consideration to the flat $D=11$ target superspace.

Worldvolume superspaces of the M2 and M5-brane are parametrized by $d=p+1$ bosonic coordinates $\xi^m$ (where $p=2$ for M2 and $p=5$ for M5) and have 16 fermionic directions associated with maximal $n=16$ supersymmetries of the $(p+1)$-dimensional superworldvolume geometry. The number of worldvolume supersymmetries is taken to be the same as the number of $\kappa$-symmetries in the Green-Schwarz-like formulations of these branes \cite{Bergshoeff:1987cm,Bandos:1997ui,Aganagic:1997zq}. The corresponding fermionic coordinates $\eta^{\mu q}$ carry two indices, the index $\mu$ which in the case of flat superspace (or after fixing a Wess-Zumino gauge in supergrvaity superspace) is associated with the index $\alpha$ of a fundamental representation of ${\tt Spin}(1,p)$, and the index $q$ which labels a fundamental representation of ${\tt Spin}(10-p)$. Note that the group ${\tt Spin}(1,p)\times {\tt Spin}(10-p)$ is the subgroup of the $D=11$ tangent space symmetry group ${\tt Spin}(1,10)$ which remains unbroken when the M2 or M5-brane worldvolume is embedded into $D=11$ target superspace. In summary, the M2- and M5-brane superworldvolume coordinates are
$$
z^M=(\xi^m,\eta^{\mu q})\,,
$$
and the superworldvolume geometry is described by the supervielbeins
\begin{eqnarray}
\label{eA=ea+} e^A= dz^{M} e_{M}{}^{A}(z) =
(e^a\; , \; e^{\alpha q}) \; , \qquad p=2,5\,,\qquad \begin{cases}m,a=0,1,\ldots , p\; , \qquad \cr
\mu,\alpha=1,\ldots, 2^{[\frac p2]} \; , \cr q=1,\ldots , 2^{[\frac {10-p}2]} \; , \end{cases}
\end{eqnarray}
where $[\frac n2]$ stands for the integer part of $\frac n2$.

The superwolrdvolume pullback of the target-space supervielbeins \eqref{11dsv} in the basis of \eqref{eA=ea+} is
\begin{equation}\label{pullbacked11sv}
E^{\underline a}(Z(z))=e^{a}E_a^{\underline a}+e^{\alpha q}E_{\alpha q}^{\underline a},\qquad E^{\underline \alpha }(Z(z))=e^{a}E_a^{\underline \alpha}+e^{\alpha q}E_{\alpha q}^{\underline \alpha},
\end{equation}
where
\begin{equation}\label{pullcoef}
E_A^{\underline A}=\nabla_A Z^{\underline M}E_{\underline M}^{\underline A}\,,\qquad
\nabla_{A}:=e_{A}^{M}(z) \partial_{M}\,
\end{equation}
and $e_{A}^{M}(z)$ are the components of the matrix inverse to the matrix $e^A_M$ of the worldvolume supervielbein components \eqref{eA=ea+}.

The superembedding condition is the vanishing of the component of the pullback of $E^{\underline a}$ in \eqref{pullbacked11sv} along the worldvolue fermionic directions $e^{\alpha q}$, i.e.
\begin{equation}
\label{SembEq}
 {E}_{\alpha q}{}^{\underline{a}}= \nabla_{\alpha q}
 {Z}^{\underline{M}}\,
E_{\underline{M}}{}^{\underline{a}}({Z}(z)) =0 \;\,.
\end{equation}
In  the case of superparticles and superstrings the geometry of the worldline superspace and the worldsheet superspace is superconformally flat which allowed us to consider the superembedding equation with flat fermionic covariant derivatives $\nabla_{\alpha q}=D_{\alpha q}$, see \eqref{Dq=n8} and \eqref{D+q=}. This is not the case for the higher-dimensional super-p-branes, in particular for the M2-brane and the M5-brane. So, in these cases the construction is invariant under the full group of worldvolume superdiffeomorphisms $z^M \to z^{'M}(z)$.

There are to ways to proceed in these cases. The first is to assume some constraints (similar to \eqref{Ta}) on the worldsheet superspace supergravity described by the supervielbeins \eqref{eA=ea+}. The second one is to induce the worldvolume superspace supergravity by a specific embedding of the superworldvolume into the target superspace. We will proceed along this way.
It requires the introduction of auxiliary superworldvolume variables which have the meaning of $SO(1,10)$ Lorentz harmonics or so called (spinor) moving frame variables.

\subsection{Superembedding condition and the induced geometry of the superworldvolume}

The superembedding condition \eqref{SembEq} can be obtained by specifying the geometry of the superwordvolume $\mathcal M_{sw}$ induced by its embedding into the target superspace as follows.
 We {\tt require} that, as in the  classical `bosonic' surface theory, the embedding is such that at each point $z^M$ in $\mathcal M_{sw}$ the pullback of the target space supervielbein
\begin{equation}\label{pbvs}
E^{\underline a}(Z(z))=dz^M\partial_MZ^{\underline M}E_{\underline M}^{\underline a}(Z(z))
\end{equation}
can be brought, by a local $SO(1,10)$ Lorentz transformation $E^{\underline a}\to E^{\underline a}u_{\underline a}{}^{\underline b}$  with a matrix
\footnote{Notice the different nature of the lower  and the upper indices of the components of the
matrix $u_{\underline b}{}^{\underline a}$. The lower index is associated with the vector representation of the $SO(1,10)$ Lorentz group, while the upper index is the cumulative index of the direct product of the vector representations of
$SO(1,p)\times SO(D-p-1)$. To simplify the presentation we will not make a notational distinction between them, and the meaning of the indices will be clear from the context.}
\begin{equation}\label{uaui}
u_{\underline a}{}^{\underline b}(z)=(u_{\underline a}{}^{a}(z), u_{\underline a}{}^{i}(z))\, \qquad a=0,1\ldots,p;\qquad i=1,\ldots, 10-p\,,
\end{equation}
to the frame in which the components of $E^{\underline a}$ orthogonal to the tangent space of $\mathcal M_{sw}$ are zero
\begin{equation}
\label{Ei=0} {E}^{i}:= {E}^{\underline{a}}
u_{\underline{a}}{}^i =0\; ,
\end{equation}
while the components of $E^{\underline a}$ along the tangent space of $\mathcal M_{sw}$ are identified with the worldvolume vector supervielbein $e^a(z)$
\begin{eqnarray}\label{Eua=ea}
{E}^a:= {E}^{\underline{b}} u_{\underline{b}}{}^{ a} = e^a(z)
\; .  \qquad
\end{eqnarray}

Performing the inverse Lorentz transformation of  \eqref{Ei=0} and  \eqref{Eua=ea} with the matrix\footnote{In \eqref{u-1}, the definition of the components of $u_{\;\;\underline b}^{-1\underline a}$, reflects the orthogonality properties of the Lorentz matrices \eqref{u}, and $\eta_{ab}$ and $-\delta_{ij}$ are the components of the Minkowski metric $\eta_{\underline{ab}}=(\eta_{ab},-\delta_{ij})$ with the almost minus signature.}
\begin{equation}\label{u-1}
u_{\;\;\underline b}^{-1\underline a}(z)=(u_{a}{}^{\underline a},u_{i}{}^{\underline a})\, \qquad
u_{a}{}^{\underline a}:=\eta_{ab}\eta^{\underline {ab}}u_{\underline{b}}{}^{b}(z)\,, \qquad
u_{i}{}^{\underline a}:=-\delta_{ij}\eta^{\underline {ab}}u_{\underline{b}}{}^{j}(z)
\end{equation}
we get back $E^{\underline a}$ in the worldvolume basis associated with the induced supervielbein $e^a$
\begin{equation}\label{pbvs1}
E^{\underline a}(Z(z))=e^a\,u_a{}^{\underline a}\,.
\end{equation}
Comparing this expression with \eqref{pullbacked11sv} we see that for the embedding of $\mathcal M_{sw}$ defined by \eqref{Ei=0} and \eqref{Eua=ea}  the superembedding condition \eqref{SembEq} holds \footnote{Note that in the case of the embedding of the pure bosonic surfaces into the target spaces the embedding conditions \eqref{Ei=0} and \eqref{pbvs1} do not impose any restrictions on the induced worldvolume geometry, since they are purely conventional.  Nevertheless, as it was discussed in \cite{Bandos:1995sy}, equation  \eqref{Ei=0}
serves as a convenient starting point to reproduce the   classical surface theory based on extrinsic geometry concepts and  a related  embedding approach to the bosonic strings  \cite{Lund:1976ze,Omnes:1977km,Barbashov:1990ce,Zheltukhin:1982gd,Bandos:1995zw}.}, and moreover the component $E_a^{\underline a}(Z(z))$ is part of the Lorentz matrix \eqref{u-1}
\begin{equation}\label{E=u}
E_a^{\underline a}=u_{a}{}^{\underline a}(z)\,.
\end{equation}
We will call the components of the auxiliary worldvolume matrix superfields \eqref{uaui} and their inverse \eqref{u-1} the Lorentz vector harmonics. Note that the conditions \eqref{Ei=0} and \eqref{Eua=ea} remain valid under the following local $SO(1,p)\times SO(10-p)$ transformations of $(u_{\underline{a}}{}^a, u_{\underline{a}}{}^i)$ and $e^a$
\begin{equation}\label{localo}
(u'_{\underline{a}}{}^a, u'_{\underline{a}}{}^i) = \left(u_{\underline{a}}{}^{b}{\mathcal O}_{b}{}^a(z), u_{\underline{a}}{}^{j}{\mathcal O}_{j}{}^i(z)\right), \qquad e'^{a}=e^{b}{\mathcal O}_{b}{}^a(z)\,.
\end{equation}
 $SO(1,p)\times SO(10-p)$ will be a local worldvolume symmetry of our construction. Since the $SO(1,10)$ Lorentz harmonics are defined modulo this symmetry, they parametrize a coset space $\frac {SO(1,D-1)}{SO(1,p)\times SO(D-p-1)}$, which explains their name\footnote{This terminology stems from the seminal papers \cite{Galperin:1984av,Galperin:1985bj} where the $SU(2)/U(1)$ and $SU(3)/[U(1)\times U(1)]$ harmonic superspaces were introduces
and used to construct, for the first time,
the off-shell superfield description of the
$N=2$ and $N=3$, $D=4$ non-Abelian supersymmetric Yang-Mills theories
(see also the book \cite{Galperin:2007wpa}). The vector Lorentz harmonics (which can be also called moving frame variables, see below) were introduced in \cite{Sokatchev:1985tc} under the name of light-cone harmonics and used their and in \cite{Sokatchev:1987nk} to describe a superparticle model.}.

Another name used for the vector Lorentz harmonics is moving frame variables. The reasons are that the matrix \eqref{uaui} describes Lorentz transformations and hence is composed of the orthogonal and normalized vectors \eqref{u}, which have the property to be orthogonal and parallel to the bosonic directions of the worldvolume superspace \begin{equation}\label{uaui=0}
u^{a\underline c}(z)u_{\underline c}{}^{b}(z)=\eta^{ab}\; ,\qquad
u^{a\underline c}(z)u_{\underline c}{}^{i}(z)=0\; ,\qquad
u^{i\underline c}(z)u_{\underline c}{}^{j}(z)=-\delta^{ij}\;.
\end{equation}
 The Lorentz frame which they define changes from `point' to `point' of the worldsheet superspace in such a way that the above properties are preserved. In the pure bosonic limit we can speak about the p-brane surface which moves in time thus forming the worldvolume, and the frame moves together with the p-brane preserving its orientation with respect to the worldvolume. Hence the name the moving frame variables.

 Let us now proceed with the definition of the induced fermionic supervielbein in $\mathcal M_{sw}$. To this end, let us note that as in the case of (the pullback of) the vector supervielbein \eqref{pbvs}, the fermionic supervielbein $E^{\underline \alpha}$ and its pullback are defined modulo a local transformation acting as the Majorana spinor representation of $SO(1,10)$ (or ${\tt Spin}(1,10)$). The parameters of this transformation form a $32\times 32$ matrix $v_{\underline \beta}{}^{\underline \alpha}$ which is related to the  matrix  $u_{\underline b}{}^{\underline a}$, eq. \eqref{uaui}, in the vector representation of $SO(1,10)$ as follows
 \begin{eqnarray}\label{VGVT=uHa}
v\Gamma^{\underline{a}}v^T = \Gamma^{\underline{b}}
u_{\underline{b}}{}^{\underline{a}} \qquad \Rightarrow \qquad \begin{cases}v\Gamma^{a}v^T =
\Gamma^{\underline{b}} u_{\underline{b}}{}^{a}\; , \cr v\Gamma^{i}v^T =
\Gamma^{\underline{b}} u_{\underline{b}}{}^{i}\end{cases}
 \;  . \qquad
\end{eqnarray}
The relation of $v_{\underline \beta}{}^{\underline \alpha}$ to the inverse matrix $u^{-1\underline a}_{\underline b}$  \eqref{u-1} is
\begin{eqnarray}\label{VTGV=uHa}
v^T{\Gamma}^{\underline{a}}v = {\Gamma}^{\underline{b}}
u_{\underline{b}}^{-1\underline{a}}= {\Gamma}^{b} u_{b}{}^{\underline{a}} + {\Gamma}^{i}
u_i{}^{\underline{a}}\; .   \qquad
\end{eqnarray}
The Lorentz group transformations also preserve the charge conjugation matrix $C_{\underline{\alpha}\underline{\beta}}$ and its inverse $C^{\underline{\alpha}\underline{\beta}}$. So the matrix  $v_{\underline \beta}{}^{\underline \alpha}$ satisfies the conditions
\begin{eqnarray}\label{VTCV=C}
vCv^T = C\; ,    \qquad v^TC^{-1}v = C^{-1}\; .    \qquad
\end{eqnarray}
These relations imply that the inverse matrix $v_{\;\;\underline \beta}^{-1\underline\alpha}$ is related to $v_{\underline \alpha}{}^{\underline\beta}$ by a kind of  transposition. The explicit expressions of the Dirac matrices and the charge conjugation matrix are given in the Appendices A and B.

So, the matrices $v$ and $u$ carry the same number of independent components which is equal to the dimension of $SO(1,10)$. This number is further reduced by the action of the gauge group $SO(1,p) \times SO(10-p)$. We will, therefore, call $v_{\underline \alpha}{}^{\underline\beta}$ the Lorentz spinor harmonics as in \cite{Bandos:1990ji,Galperin:1991gk,Delduc:1991ir,Bandos:1992hu,Bandos:1993yc,Bandos:1994eu,Bandos:1995zw}.
 Another name, spinor moving frame variables, used for the spinor Lorentz harmonics in \cite{Bandos:1992np,Bandos:1992hu,Bandos:1993yc,Bandos:2017zap,Bandos:2017eof}, reflects the fact that they provide a kind of  square root of the moving frame vectors, i.e. the vector Lorentz harmonics, in the sense of equations \eqref{VGVT=uHa} and \eqref{VTGV=uHa}.
In other words they are counterparts of  $D=4$ diades which are used to construct a light-like tetrad of the Newmann-Penrose formalism of General Relativity \cite{Newman:1961qr} by using the Cartan-Penrose representation for the light-like 4-vector \cite{Penrose:1972ia}.

As in the case of the vector supervielbein, we can use $v_{\underline \alpha}{}^{\underline\beta}$ to associate certain components of the pullback of the spinor supervielbein
\begin{equation}\label{spinortsv}
E^{\underline \alpha}=dz^M\partial_MZ^{\underline M}E_M^{\underline \alpha}{{(Z(z))}}=e^aE_a^{\underline\alpha}+e^{\alpha q}E_{\alpha q}^{\underline\alpha}
\end{equation}
with the induced spinor supervielbein $e^{\alpha q}$ in the superworldvolume. To this end, as in \eqref{uaui}, we split the index $\underline\alpha$ into two 16-dimensional sets of indices, one of which is associated with the worldvolume spinor indices $\alpha q$, and another one labelling a complimentary spinor representation which depends on the dimension of the superworldvolume, namely
\begin{eqnarray}\label{VharmM2} v_{\underline{\beta}}{}^{\underline{\alpha}}= \left(
v^{\;\;\alpha q}_{\underline{\beta}},~ v_{\underline{\beta}\, \alpha \dot q} \right)\;
\in \; Spin(1,10) \; , \qquad \begin{cases} \alpha ,\beta =1,2 \; , \cr q,\dot q=1,
\ldots , 8\; ,\end{cases} \quad for \;\; M2-brane \; , \quad
\\ \label{VharmM5} v_{\underline{\beta}}{}^{\underline{\alpha}}= \left(
v^{\;\;\alpha q}_{\underline{\beta}},~ v_{\underline{\beta}}{}_{\alpha}^{\; q} \right)\;
\in \; Spin(1,10) \; , \quad \begin{cases} \alpha ,\beta =1,2,3,4 \; , \cr q=1,2,3,4
\; , \end{cases}  \quad for \;\; M5-brane \; . \quad
\end{eqnarray}
We identify the projection of the pullback of \eqref{spinortsv} along $v^{\;\;\alpha q}_{\underline{\beta}}$  with  the (induced) worldvolume spinor supervielbein
\begin{eqnarray}\label{e=EV}
e^{\alpha q}= E^{\underline{\beta}}v_{\underline{\beta}}{}^{\alpha q}\; .  \qquad
\end{eqnarray}
Note that with such an identification
\begin{equation}\label{datv}
E_a^{\underline \alpha}v_{\underline \alpha}{}^{\alpha q}=0.
\end{equation}

Thus the introduction of the Lorentz harmonic variables allowed us to construct the basic objects, i.e. the supervielbeins \eqref{Eua=ea} and \eqref{e=EV} describing an induced geometry of the worldvolume superspace, and also to reformulate the superembedding condition \eqref{SembEq} in the alternative form given by \eqref{Ei=0} and \eqref{Eua=ea}.

\subsection{Induced $SO(1,p)\times SO(10-p)$ connections on $\mathcal M_{sw}$}
We shall now study the consequences of the embedding conditions \eqref{Ei=0} and \eqref{Eua=ea} by taking their external differential
\begin{eqnarray}
\label{dEi=0} d{E}^i=dE^{\underline{a}}\, u_{\underline{a}}{}^i +E^{\underline{a}}\, du_{\underline{a}}{}^i =0\; ,
\end{eqnarray}
\begin{equation}\label{dea}
dE^{a}=dE^{\underline a}\,u_{\underline{a}}{}^a+E^{\underline a}du_{\underline{a}}{}^a\,.
\end{equation}
This will lead us to the definition of the  $SO(1,p)\times SO(10-p)$ connections of the induced superworldvolume geometry.

In the flat target superspace, which we further assume, the supervielbeins are the Volkov-Akulov one-form \eqref{VA1f} and $E^{\underline \alpha}=d\Theta^{\underline \alpha}$, and $dE^{\underline a}$ is given in \eqref{Taflat}. Then \eqref{dEi=0} takes the form
\begin{eqnarray}
\label{dEi=01} d{E}^i=-i d\Theta^{\underline{\alpha}}
{\Gamma}^{\underline{a}}_{\underline{\alpha}\underline{\beta}} d\Theta^{\underline{\beta}}\, u_{\underline{a}}{}^i +E^{\underline{a}}\, du_{\underline{a}}{}^i =0\; .
\end{eqnarray}
Let us now elaborate on the second term $E^{\underline{a}}\, du_{\underline{a}}{}^i$ of this equation. Using the orthogonality properties \eqref{u} of the Lorentz harmonics \eqref{uaui} and \eqref{u-1}, and the embedding condition \eqref{Eua=ea}, we have
\begin{equation}\label{Edu}
E^{\underline{a}}\, du_{\underline{a}}{}^i=E^{\underline{b}}\, u_{\underline b}{}^{\underline c}\,u^{-1\underline a}_{\underline c}du_{\underline{a}}{}^i=e^a\,u_a{}^{\underline a}du_{\underline{a}}{}^i+E^j\,u_j{}^{\underline a}du_{\underline{a}}{}^i\,,
\end{equation}
where the last term is actually zero due to the embedding condition \eqref{Ei=0}.

We now notice that the quantities $u_a{}^{\underline a}du_{\underline{a}}{}^i$ and $u_j{}^{\underline a}du_{\underline{a}}{}^i$ in \eqref{Edu}
are nothing but components of the $SO(1,10)$ Cartan form \eqref{Omegaflat}
 \begin{eqnarray}\label{Omegaia}
\Omega_a{}^{i}= u_a{}^{\underline a}du_{\underline{a}}{}^i=-(dE_a^{\underline a})u_{\underline a}{}^i\,,\qquad \\  \label{Omegaij} \Omega_j{}^{i}= u_j{}^{\underline a}du_{\underline{a}}{}^i\;.\qquad
\end{eqnarray}
(remember the relation \eqref{E=u}).

The one-form $\Omega_j{}^{i}$ takes values in the $SO(10-p)$ subgroup of $SO(1,10)$ and is associated with the connection of the $SO(10-p)$ gauge symmetry in $\mathcal M_{sw}$. We thus rewrite \eqref{dEi=01} as follows
\begin{eqnarray}
\label{dEi=02} \mathcal D{E}^i:=dE^i-E^j\Omega_j{}^i= e^a\,\Omega_a{}^i-i d\Theta^{\underline{\alpha}}
{\Gamma}^{\underline{a}}_{\underline{\alpha}\underline{\beta}} d\Theta^{\underline{\beta}}\, u_{\underline{a}}{}^i=0\; ,
\end{eqnarray}
where $\mathcal D$ is the  covariant derivative which includes the connection $\Omega_j{}^i$ when it acts on the quantities that carry $SO(10-p)$ indices.

The decomposition of the form $\Omega_a{}^{i}$ in the basis of the worldvolume supervielbeins is
\begin{eqnarray}\label{Ombi=Mp}
\Omega_a{}^{i}=e^{\alpha q} \Omega_{\alpha q\,a}{}^{i}+ e^b\Omega_{ba}{}^{i}\,.
\end{eqnarray}
We see that the symmetric tensor component $\Omega_{(ba)}{}^i$ of the form $\Omega_{a}{}^i$  does not enter the relation \eqref{dEi=02} and hence remains arbitrary at this stage. The explicit expression for this component is
\begin{eqnarray}\label{Kabi=M2}
 K_{a b}{}^i:=\Omega_{(a\, b)}{}^i= - {\nabla}_{(a}{E}_{b)}{}^{\!\underline{\,c}}\; u_{\underline{c}}{}^i\;
\qquad
\end{eqnarray}
 One can recognize in $K_{a b}{}^i(z)$ a  superfield generalization of the second fundamental form of the classical surface theory. In our context it characterizes the worldvolume superspace considered as a supersurface in the $D=11$ target
superspace. The dynamical equations of motion of the M2 and M5 branes will arise as algebraic conditions
on the leading ($\eta =0$) component of
this superfield.

Let us elaborate on the form of the differential \eqref{dea} of the embedding condition \eqref{Eua=ea} in the same way as we did for \eqref{dEi=0}. Thus, using \eqref{Ei=0} we get
\begin{equation}\label{dea=}
de^a=-i d\Theta^{\underline{\alpha}}
{\Gamma}^{\underline{a}}_{\underline{\alpha}\underline{\beta}} d\Theta^{\underline{\beta}}\, u_{\underline{a}}{}^a +e^{b}\, u_b{}^{\underline a}du_{\underline{a}}{}^a,
\end{equation}
where $u_b{}^{\underline a}du_{\underline{a}}{}^a$ can be identified with the induced $SO(1,p)$ spin connection in $\mathcal M_{sw}$
\begin{equation}\label{spin1p}
\Omega_{b}{}^a=u_b{}^{\underline a}du_{\underline{a}}{}^a\,.
\end{equation}
Then \eqref{dea} takes the form of a constraint on the vector component of the torsion of $\mathcal M_{sw}$
\begin{equation}\label{dea1}
T^a={\mathcal D}e^a:=de^a-e^b\,\Omega_b{}^a=-i d\Theta^{\underline{\alpha}}
{\Gamma}^{\underline{a}}_{\underline{\alpha}\underline{\beta}} d\Theta^{\underline{\beta}}\, u_{\underline{a}}{}^a\,.
\end{equation}
On the examples of the M2- and M5-brane we will see that \eqref{dea1} is one of the essential constraints of the supergravity geometry on $\mathcal M_{sw}$.

To proceed further we should also define superworldvolume covariant derivatives of the Lorentz spinor harmonics $v_{\underline \beta}{}^{\underline \alpha}$. In view of their relation to the Lorentz vector harmonics \eqref{VGVT=uHa} and \eqref{VTGV=uHa}, the corresponding Cartan forms are related as follows
\begin{eqnarray}\label{CFspin=CFso}
v^{-1}dv &=&  {1\over 4}
(u^{-1}du)^{\underline{a}\underline{b}} \Gamma_{\underline{a}\underline{b}}={1\over 4} \Omega^{\underline{a}\underline{b}}
\Gamma_{\underline{a}\underline{b}}
\qquad
\nonumber \\
 &=& {1\over 4} \Omega^{ab}
\Gamma_{ab} + {1\over 4} \Omega^{ij} \Gamma^{ij} + {1\over 2} \Omega^{ai}
\Gamma_a\Gamma_i \; .   \qquad
\end{eqnarray}
Using this relation one defines the ${\tt Spin}(1,p)\times {\tt Spin}(D-p-1)$-covariant derivative acting on the upper indices of $v_{\underline\beta}{}^{\underline \alpha}$ split into ${\tt Spin}(1,p)\times {\tt Spin}(D-p-1)$ indeces as in \eqref{VharmM2} and \eqref{VharmM5}
\begin{eqnarray}\label{DV=VOm}
{\cal D}v &:=& dv  - {1\over 4} v\Gamma_{ab}\, \Omega^{ab} -
{1\over 4} v\Gamma^{ij} \, \Omega^{ij} =  {1\over 2} v\Gamma_a\Gamma_i\, \Omega^{ai} \;
 \qquad
\end{eqnarray}
Note that  the covariant derivative of the Lorentz spinor harmonics is expressed in terms of the covariant Cartan form
\eqref{Omegaia}.
To write an explicit form of \eqref{DV=VOm}  we  need an $SO(1,p) \times
SO(D-p-1)$ invariant representation for the matrices
$\Gamma^{\underline{a}} = ( \Gamma^{a}, \Gamma^{i} )$.
which is $p$--dependent. For the M2- and M5-brane cases they are given in Appendices A and B.

\subsection{M2-brane} \label{M2SEmb}

We will now show how the superembedding conditions produce the dynamical equations of motion of the M2-brane. Technical material on which the discussion is based is collected in Appendix A.

The M2-brane superworldvolume is an $n=8$,  $d=3$ superspace $\mathcal M^{(3|16)}_{sw}$, where $n$ stands for the number of two-component Majorana spinors in $d=2+1$ so that the total number of the superworldvolume fermionic directions is 16. The  geometry of  $\mathcal M^{(3|16)}_{sw}$   induced by its embedding into flat $D=11$ superspace is characterized by the worldvolume supervielbeins determined by the superembedding conditions \eqref{Ei=0}, \eqref{Eua=ea} and \eqref{e=EV}, and  by the corresponding connections of the local $SO(1,2)\times SO(8)$ symmetry \eqref{Omegaij} and \eqref{spin1p}. By construction the  $\mathcal M^{(3|16)}_{sw}$    geometry is that of an  $n=8$, $d=3$ supergravity characterized by the torsion constraint \eqref{dea1} the r.h.s. of which is still to be elaborated.

Let us now have a look at the structure of the projection of the pullback of the target-space spinorial supervielbein $E^{\underline\alpha}=d\Theta^{\underline \alpha}$ along the fermionic directions `orthogonal' to those of $\mathcal M^{(3|16)}_{sw}$. This projection is made by the Lorentz spinor harmonics $v_{\underline \alpha\,\alpha \dot q}$ \eqref{VharmM2}\footnote{Note that the index $\dot q=1,\ldots,8$  labels a spinor representation of $SO(8)$ which is different from the $SO(8)$ spinor representation labelled by the index $q=1,\ldots,8$ in \eqref{e=EV} and from the $SO(8)$ vector representation labelled by the index $i=1,\ldots,8$ in \eqref{Ei=0}. The three representations are related by triality manifested in the $SO(8)$ invariance of the matrices $\gamma^i_{q\dot q}$. See Appendix A for more details.}. In the worldvolume supervielbein basis \eqref{Eua=ea} and \eqref{e=EV} we have
\begin{eqnarray}\label{Ef2=M20}
E_{\alpha\dot q}:= d\Theta^{\underline{\beta}}\,
v_{\underline{\beta}\alpha\dot q} =
e^{\beta r} h_{\beta r\; \alpha\dot q}+ e^b \psi_{b\; \alpha\dot q}
\; , \qquad
\end{eqnarray}
where $h_{\alpha q\; \beta \dot q}(z)$ and $\psi_{b\; \beta\dot q}(z)$ are bosonic and fermionic superfields in $\mathcal M^{(3,16)}_{sw}$.

To find constraints on these superfields let us elaborate on the consequence \eqref{dEi=02} of the superembedding condition \eqref{Ei=0}. Using the decompositions \eqref{Ombi=Mp} and \eqref{Ef2=M20}, and the relation \eqref{11DuiG=AP} between the spinor and vector Lorentz harmonics, we write \eqref{dEi=02} as follows
\begin{equation}\label{eOai}
e^ae^{\alpha q} \,\Omega_{\alpha q\,a}{}^{i}+ e^ae^b\,\Omega_{ba}{}^{i}=-2ie^{\alpha q}\gamma^i_{q\dot q}E_{\alpha \dot q}=-2ie^{\beta r}e^{\alpha q}\gamma^i_{q\dot q}h_{\beta r\; \alpha \dot q}+2ie^be^{\alpha q}\gamma^i_{q\dot q}\psi_{b\; \alpha\dot q},
\end{equation}
where $\gamma^i_{q\dot q}$ are $SO(8)$ counterparts of the Pauli matrices.

Comparing the components of the left- and the right-hand side of the above equation we see that
\begin{equation}\label{Oi[ab]}
\Omega_{[ab]}^i=0, \qquad \Omega_{\alpha q\,a}{}^{i}=2i
\gamma^i_{q\dot{p}} \psi^a_{\; \alpha \dot{p}}\,
\end{equation}
and
\begin{equation}\label{h+h}
\gamma^i_{q\dot{q}}  h_{\beta r\,\alpha \dot q}+
\gamma^i_{r\dot{q}}  h_{\alpha q\, \beta \dot q}=0\,,
\end{equation}
whose solution is trivial
$$
h_{\beta r\,\alpha \dot q}=0.
$$
So,  (\ref{Ef2=M20}) simplifies to
\begin{eqnarray}\label{Ef2=M2}
{E}_{\alpha \dot q}:=d\Theta^{\underline\beta}v_{\underline\beta\,\alpha \dot q}=
 e^b \psi_{b\; \alpha\dot q} \;,
\end{eqnarray}
while, in view of \eqref{Oi[ab]}, the covariant form \eqref{Ombi=Mp} reduces to
\begin{equation}\label{Omai=M2}
 \Omega_a{}^{i}=   2i e^{\alpha q}
\gamma^i_{q\dot{q}}\psi_{a\,\alpha \dot{q}} + e^b  K_{ab}^i,
\end{equation}
where the symmetric tensor $K_{ab}^i$ was identified as the second fundamental form in \eqref{Kabi=M2}, and
\begin{eqnarray}\label{chidef}
 \psi_{a\,\alpha\dot{q}} = {E}_b^{\underline{\alpha}}\,
v_{\underline{\alpha}\,\alpha\dot{q}}= \nabla_b\Theta^{\underline{\alpha}}\,
v_{\underline{\alpha}\,\alpha\dot{q}}\;  .
\end{eqnarray}

Using the equation (\ref{Ef2=M2}) and   \eqref{11DuaG=AP} of Appendix A we can
find the explicit form of the $\mathcal M^{(3,16)}_{sw}$ torsion \eqref{dea}
\begin{eqnarray}\label{Dea=M2}
T^a={\cal D}e^a= {\textcolor{blue}{-}}i e^{\alpha q} \, e^{\beta q}\gamma^a_{\alpha\beta} - i
e^b\,e^c \, \psi_{b\alpha \dot q}\gamma^{a\alpha\beta} \psi_{c\beta\dot r}\delta^{\dot q\dot r}\; , \qquad
\end{eqnarray}
 which contains the basic essential constraint $T_{\alpha\beta}^a ={\textcolor{blue}{-}}2i \gamma^{a}_{\alpha\beta}$ of the supergravity theory.

Now let us take the worldvolume covariant derivative of the left- and the right-hand side of \eqref{Ef2=M2}. The derivative (acting from the right) on the left-hand side reads
\begin{eqnarray}\label{Dlhs}
\mathcal D(d\Theta^{\underline\beta}v_{\underline\beta\,\alpha \dot q})&=&d\Theta^{\underline\beta}\mathcal Dv_{\underline\beta\,\alpha \dot q}=- {1\over 2} e^{\beta p} \,\Omega_a{}^{i}\, \gamma^a_{\alpha\beta}
 \gamma^i_{p\dot{q}} \nonumber\\
 &=& (i e^{\beta p} \,e^{\delta q}
\gamma^i_{q\dot{q}}\psi_{a\,\delta \dot{q}}-\frac 12 e^{\beta p} e^a K_{ab}^i)  \gamma^a_{\alpha\beta}\gamma^i_{p\dot{q}}
\end{eqnarray}
where we used that $e^{\alpha q}=d\Theta^{\underline\alpha}v^{\alpha q}_{\underline\alpha}$ and the expression for the covariant
 derivative of the spinorial harmonics $v_{\underline{\alpha} \, \alpha \dot{q}}$ (see {\eqref{Dv=M2}}) derived from (\ref{DV=VOm}) with the use of the gamma-matrix representation given in Appendix A.
  In the last step we substituted the explicit expression \eqref{Omai=M2} for $\Omega_a{}^i$.

The covariant derivative of the right-hand side of \eqref{Ef2=M2}, with taking into account \eqref{Dea=M2}, gives
\begin{equation}\label{Drhs}
\mathcal D(e^b\psi_{b\alpha \dot q})= -2i e^{\beta p}  e^{\delta p} \gamma^b_{\beta\delta}\psi_{b\alpha
 \dot{q}}  + i e^b  e^c \,
(\psi_b\gamma^a \psi_c )\, \psi_{a \alpha\dot q} +e^b\,\mathcal D
\psi_{b \alpha\dot q}\;
\end{equation}
We should now equate the right-hand sides of \eqref{Dlhs} and \eqref{Drhs}. For the components of $e^{\beta p}\wedge e^{\delta q}$ we get
\begin{equation}\label{ff}
 \gamma^i_{p\dot{q}}
\gamma^i_{q\dot{p}} \gamma^{a}_{\alpha \beta} \psi_{a \gamma
\dot{p}}+\gamma^i_{q\dot{q}} \gamma^i_{p\dot{p}}
\gamma^{a}_{\alpha \gamma} \psi_{a \beta \dot{p}}=- 2\delta_{p\,
q} \psi_{a \, \alpha\dot{q}}\,.
\end{equation}
The only consequence of this equation is $\gamma_{\; \alpha}^{a\,\beta} \psi_{a\,\beta \dot{q}} =0$ which in view of the explicit form \eqref{chidef} of $\psi_{a\,\beta \dot{q}}$ is a first-order Dirac-like differential equation for the superfield $\Theta^{\underline \alpha}$
\begin{eqnarray}\label{fEqmM2}
{\gamma}_{\;\alpha}^{a\,\beta} \nabla_a\Theta^{\underline{\alpha}}
v_{\underline{\alpha}\beta \dot{q}}=0 \;  .
\end{eqnarray}
At $\eta=0$ this equation coincides with the physical equations  of motion of the fermionic embedding coordinates $\theta^{\mathcal \alpha}(\xi)$ of the M2-brane, while at higher orders it does not produce any further restrictions on the independent dynamical variables of M2.

To obtain the bosonic field equations of the M2-brane let us equate the components of $e^{b}\wedge e^{\beta p}$ in \eqref{Dlhs} and \eqref{Drhs}. We thus get
\begin{equation}\label{bf}
 {1\over 2}\gamma^i_{p\dot{q}}
\gamma^a_{\alpha\beta} K_{ab}{}^i=\mathcal D_{\beta p} \psi_{b \alpha \dot{q}}.
\end{equation}
Contracting this equation with
${\gamma}^{b \gamma\alpha}$ one gets
\begin{eqnarray}\label{giKaai=}
\gamma^i_{p\dot{q}} K_{a}{}^{ai} \delta_\beta{}^\gamma =  2 {\cal D}_{\beta p}({\gamma}^{a } \psi_{a
\dot{q}})^\gamma \; . \qquad
\end{eqnarray}
The right-hand side of the above equation vanishes due to the fermionic equation of motion
(\ref{fEqmM2}), and we arrive at the bosonic equations of motion of the M2-brane
\begin{eqnarray}\label{bEqmM2}
 K_{a}{}^{ai} := -\nabla^a {E}_a{}^{\underline{b}}\; u_{\underline{b}}{}^i = 0, \qquad E_a^{\underline b}=\nabla_aX^{\underline b}-i\nabla_a\Theta\Gamma^{\underline a}\Theta\,.
\end{eqnarray}
To demonstrate a contact with the standard (Green-Schwarz-like) formulation of the supermembrane
\cite{Bergshoeff:1987cm,Bergshoeff:1987qx}, let us consider the leading $\eta=0$ component of \eqref{bEqmM2}  and  formally set to zero the fermionic components of the supervielbein $e_a^M$, in particular $e_a^\mu\vert_{\eta=0}=0$.
Then \eqref{bEqmM2} reduces to differential equations on ordinary worldvolume fields $x^{\underline a}(\xi)$ and $\theta^{\underline\alpha}(\xi)$ which can be rewritten with the use of the induced worldvolume metric  $g_{mn}(\xi)=e_m{}^ae_{an}=
{E}_m{}^{\underline{a}}{E}_{n\underline{a}}$ (where now $E^{\underline a}_m=\partial_mx^{\underline a}-i\partial_m\theta\Gamma^{\underline a}\theta$) as follows
\begin{eqnarray}\label{bEqM2=st}
\nabla^a {E}_a{}^{\underline{b}}\; u_{\underline{b}}{}^i=\frac 1{\sqrt{|g|}}\partial_m (\sqrt{|g|}g^{mn}{E}_n{}^{\underline{b}})u_{\underline{b}}{}^i= 0\,.
\end{eqnarray}
The equation \eqref{bEqM2=st} is the projection along $u_{\underline a}{}^i(\xi)$ of the GS supermembrane equation of motion
\begin{equation}\label{gsmeq}
\partial_m (\sqrt{|g|}g^{mn}{E}_n{}^{\underline{b}})=0.
\end{equation}
The other projection of this equation  which is tangential to the worldvolume, \linebreak
$\partial_m (\sqrt{|g|}g^{mn}{E}_n{}^{\underline{b}}) u_{\underline{b}}{}^b = 0,$
or equivalently
\begin{equation}\label{uapro}
\partial_m (\sqrt{|g|}g^{mn}{E}_n{}^{\underline{b}}) u_{\underline{b}}{}^be_{b\,l} = 0\,,
\end{equation}
is satisfied identically. This is a Noether identity reflecting the reparametrization invariance of the supermembrane action. Indeed, remembering that $u_{a}{}^{\underline a}=E^{\underline a}_a$ (see  eq. \eqref{E=u}) and that $u_{\underline a}{}^a=\eta^{ab}\eta_{\underline a\underline b}u_{b}{}^{\underline b}$, we have $u_{\underline b}{}^{b}e_{bl}=E_{l}^{\underline a}\eta_{\underline a\underline b}$. So \eqref{uapro} takes the form
\begin{equation}\label{uapro1}
\partial_m (\sqrt{|g|}g^{mn}{E}_n{}^{\underline{b}}) E^{\underline a}_l \eta_{\underline a\underline b}= 0\,.
\end{equation}
Integrating this equation by parts and using the torsion constraint \eqref{Taflat} we get
\begin{equation}\label{uapro2}
\partial_l \left(\sqrt{|g|}\right)-\sqrt{|g|}g^{mn}{E}_n{}^{\underline{b}}\eta_{\underline a\underline b}\partial_lE^{\underline a}_m+i\sqrt{|g|}g^{mn}\,E^{\underline a}_n\partial_m\Theta\Gamma_{\underline a} \partial_l\Theta= 0\,.
\end{equation}
Since $g_{mn}(\xi)={E}_m{}^{\underline{a}}{E}_{n\underline{a}}$, the first two terms on the left-hand side of this equation cancel each other, while the third term is proportional to the fermionic equations of motion and hence also vanishes.

Finally, let us confront the $e^{b}\, e^{c}$ components of \eqref{Dlhs} and \eqref{Drhs}. In \eqref{Dlhs} this  component  is absent, hence the corresponding component in \eqref{Drhs} mast vanish, i.e.
\begin{equation}\label{Dchi=}{\cal D}_{[c}\psi_{b]\alpha \dot q}=
-i \psi_c\tilde{\gamma}^a\psi_b\, \psi_{a\alpha \dot q}.
\end{equation}
One can check that this is indeed the case. To this end one should take the  covariant derivative of \eqref{chidef}, antisymmetrize it with respect to the indices $[a,b]$, and then use the form of the torsion \eqref{Dea=M2}, the expression of the covariant derivative of $v_{\underline \alpha \alpha \dot q}$ in \eqref{Dv=M2} and the condition \eqref{datv}.

We have thus shown that the  superembedding conditions imposed on the embedding of the $n=8$, $d=3$  worldvolume superspace into the $D=11$ superspace completely determines the on-shell dynamics of the M2-brane.

\subsubsection{Kappa-symmetry from worldvolume superdiffeomorphisms}
As we have mentioned, the superembedding construciton is invariant under the superdiffeomoprphisms of the worldvolume supercoordinates $z^{M}\to z'^{M}(Z)$. The M2-brane coordinate functions $Z^{\underline M}(z)=(X^{\underline m}(z),\Theta^{\underline \mu}(z))$ transform under these diffeomorphisms as scalar worldvolume superfields. Their infinitesimal transformations are
$$
\delta Z^{\underline M}(z)=\delta z^M\partial_MZ^{\underline M}(z)=i_\delta e^A\nabla_AZ^{\underline M}(z).
$$
We can project these transformations along the components $E_{\underline M}^{\underline A}$ of the target space supervielbein
\begin{equation}\label{deltaEa}
i_\delta E^{\underline a}=i_\delta e^A\nabla_AZ^{\underline M}E_{\underline M}^{\underline a}\,,
\end{equation}
\begin{equation}\label{deltaEal}
i_\delta E^{\underline \alpha}=i_\delta e^A\nabla_A Z^{\underline M}E_{\underline M}^{\underline \alpha}\,.
\end{equation}
Let us now restrict ourselves to the fermionic diffeomorphisms only, i.e. set $i_\delta e^a=0$ and $i_\delta e^{\alpha q}=\kappa^{\alpha q}(z)$. Then, due to the superembedding condition \eqref{SembEq}, the equations \eqref{deltaEa} and \eqref{deltaEa} reduce to
\begin{equation}\label{deltaEa1}
i_\kappa E^{\underline a}=\kappa^{\alpha q} E_{\alpha q}^{\underline a}=0,
\end{equation}
and
\begin{equation}\label{deltaEal1}
i_\kappa E^{\underline \alpha}=\kappa^{\alpha q}\nabla_{\alpha q} Z^{\underline M}E_{\underline M}^{\underline \alpha}=\kappa^{\alpha q}v_{\alpha q}{}^{\underline \alpha},
\end{equation}
where in the last equality we used the fact that $E_{\alpha q}^{\underline \alpha}=v_{\alpha q}{}^{\underline \alpha}$, as a consequence of the superembedding conditions \eqref{e=EV} and \eqref{Ef2=M2}.

Let us now note that since the parameter $\kappa^{\alpha q}(z)$ is arbitrary, we can equivalently express it as the $v_{\underline \beta}{}^{\alpha q}$ projection of a $D=11$ spinor superfield parameter $\kappa^{\underline \beta}(z)$,
$$
\kappa^{\alpha q}(z)=\kappa^{\underline \beta}(z)v_{\underline \beta}{}^{\alpha q}(z).
$$
Then the fermionic variation \eqref{deltaEal1} takes the form
\begin{equation}\label{deltaEal2}
i_\kappa E^{\underline \alpha}=\kappa^{\underline \beta}(v_{\underline \beta}{}^{\alpha q}v_{\alpha q}{}^{\underline \alpha})=\frac 12 \kappa^{\underline \beta}(1-\bar\Gamma)_{\underline \beta}{}^{\underline \alpha},
\end{equation}
where we used the Lorentz-harmonic relations \eqref{I=vv+M2} and \eqref{bG:=M2}.

At $\eta^{\alpha q}=0$, the fermionic diffeomorphism variations \eqref{deltaEa1} and \eqref{deltaEal2} are equivalent to the kappa-symmetry transformations of the M2-brane coordinate functions $x^{\underline a}(\xi)$ and $\theta^{\underline \alpha}(\xi)$ which leave invariant the M2-brane action of \cite{Bergshoeff:1987cm,Bergshoeff:1987qx}.

\subsection{M5-brane}

The on-shell dynamics of the M5-brane is also fixed by the superembedding equation (\ref{SembEq}) \cite{Howe:1996yn}, or by its equivalent form  (\ref{Ei=0}) and \eqref{Eua=ea}. It is
amazing that the superembedding condition also contains non-linear equations of motion of a chiral two-form gauge field living on the M5-brane, whose three-form field strength is self-dual. We will demonstrate this fact in this section. Some technical relations required to follow the discussion
are given in Appendix B.

The M5-brane superworldvolume is an $n=(2,0)$, $d=6$ superspace $\mathcal M^{(6|16)}_{sw}$, where $n$ here traditionally  stands for the number of eight-component  spinors in $d=5+1$ so that the total number of the superworldvolume fermionic directions is 16.  The  geometry of $\mathcal M^{(3|16)}_{sw}$ induced by its embedding into flat $D=11$ superspace is characterized by the worldvolume supervielbeins determined by the superembedding conditions \eqref{Ei=0}, \eqref{Eua=ea} and \eqref{e=EV}, and  by the corresponding connections of the local $SO(1,5)\times SO(5)$ symmetry \eqref{Omegaij} and \eqref{spin1p}. By construction the $\mathcal M^{(3|16)}_{sw}$ geometry is that of an $n=(2,0)$, $d=6$  supergravity characterized by the torsion constraint \eqref{dea1}. This supergravity is composite (induced by superembedding) and thus does not carry independent dynamical degrees of freedom.

 The projections of the pullback of the fermionic supervielbein $E^{\underline \alpha}=d\Theta^{\underline \alpha}$ on the Lorentz  harmonic superfields \eqref{VharmM5} are
\begin{eqnarray}\label{Ef1=M5}
{E}^{\alpha q}&:= & E^{\underline{\beta}}
v_{\underline{\beta}}{}^{\alpha q} =e^{\alpha q}\; , \qquad \\
\label{Ef2=M5} E_{\beta}^{q}\; &:=& E^{\underline{\alpha}}
v_{\underline{\alpha}}{}_{\beta}^{q} \; =
e^{\alpha q} h_{\alpha p\; \beta}{}^q + e^b \psi_{b\; \beta}{}^{q}
\; . \qquad
\end{eqnarray}
Note that the upper and lower spinor indices $\alpha$ of ${\tt Spin}(1,5)$ are not related to each other by a charge conjugation matrix (see Appendix B).

Using
 eqs. (\ref{Ef1=M5}) and (\ref{Ef2=M5}), and the decomposition of unity in terms of the Lorentz harmonics,  $\delta_{\underline{\beta}}{}^{\underline{\alpha}} = v_{\underline{\beta}}{}^{\alpha q} v_{\alpha
q}{}^{\underline{\alpha}} +  v_{\underline{\beta} \alpha}{}^q
v_q^{ \alpha\underline{\alpha}}$,
we get
\begin{eqnarray}\label{Ef=eV(h)+}
d\Theta^{\underline{\alpha}}=  e^{\beta q} E_{\beta q}{}^{\underline{\alpha}}  + e^a \psi_{a\beta}{}^p v_p{}^{\beta \underline{\alpha}}\; , \qquad  E_{\beta q}{}^{\underline{\alpha}} =
 v_{\beta p}{}^{\underline{\alpha}}+  h_{ \beta q\,\gamma}{}^p v_{p}{}^{\gamma \underline{\alpha}}\; ,  \qquad
\end{eqnarray}

Let us now consider the consequences of \eqref{dEi=02}. We have
$$
e^a\,\Omega_a{}^i=i d\Theta^{\underline{\alpha}}
{\Gamma}^{\underline{a}}_{\underline{\alpha}\underline{\beta}} d\Theta^{\underline{\beta}}\, u_{\underline{a}}{}^i
$$
or, explicietly in the basis of $(e^a,e^{\alpha q})$
\begin{equation}\label{eOaim5}
e^ae^{\alpha q} \,\Omega_{\alpha q\,a}{}^{i}+ e^ae^b\,\Omega_{ba}{}^{i}=-2e^{\alpha q}\gamma^i_{qr}E_{\alpha r}=-2e^{\beta r}e^{\alpha q}\gamma^i_{qp}h_{\beta r\; \alpha}{}^p-2e^be^{\alpha q}\gamma^i_{qr}\psi_{b\; \alpha}{}^r,
\end{equation}
where  to arrived at the right-hand side we used the identities \eqref{M5:uiG=} and the form of \eqref{Ef1=M5} and \eqref{Ef2=M5}.
Since the left-hand side of the above equation does not have an $e^{\beta r}e^{\alpha q}$  component, from  its right hand side we conclude that
\begin{equation}\label{ffh}
h_{\beta p\; \alpha}{}^r\gamma^i_{qr}+h_{\alpha q\; \beta}{}^r\gamma^i_{pr}=0\,.
\end{equation}
Using the properties \eqref{Cliff5d}--\eqref{so(5)id} of $\gamma^{i}_{qp}$ we find that \eqref{ffh} implies
\begin{equation}\label{hab}
h_{\alpha p\;\beta}{}^q = h_{\alpha\beta}\delta_p{}^q, \qquad h_{\alpha\beta}=h_{\beta\alpha}.
\end{equation}
In $d=6$, the basis of the covariant symmetric spin tensor matrices is formed by the antisymmetric product of three matrices $\gamma^a$ which is anti-self-dual (see \eqref{6Dgammas})
\begin{equation}\label{gabc}
\gamma^{abc}_{\alpha\beta} =\gamma^{[a}_{\alpha\gamma}\tilde\gamma^{b|\gamma\delta}\gamma^{|c]}_{\delta\alpha},\qquad \gamma^{abc}_{\alpha\beta}=- {1\over 3!}\epsilon^{abcdef}(\gamma_{def})_{\alpha\beta}.
\end{equation}
Therefore, we can express $h_{\alpha\beta}$ as follows
\begin{eqnarray}\label{M5:h=h3g3}
   h_{\alpha\beta}= {1\over 3!} h_{abc}\gamma^{abc}_{\alpha\beta}\; ,
\end{eqnarray}
where $h_{abc}$ is an antisymmetric self-dual tensor
\begin{eqnarray}\label{M5:h=*h}
   h_{abc}= {1\over 3!} \epsilon_{abcdef} h^{def}\; .  \qquad
\end{eqnarray}
A useful identity, which follows from the self-duality properties of $h^{abc}$ and gamma-matrix properties \eqref{6Dgammas}, is
\begin{eqnarray}\label{M5:htgh=}
  h_{\alpha\gamma}\tilde{\gamma}^{a\gamma\delta}h_{\delta\beta}= \gamma^b_{\alpha\beta}k_b{}^a\; , \qquad k_b{}^a= -2 h_{bcd}h^{cda}\;  .  \qquad
\end{eqnarray}

We shall now show that the self-dual tensor $h_{abc}$ is related to the field strength of a chiral two-form gauge  field
living in the M5-brane worldvolume superspace
$$
b_2 =\frac 1 2 dz^N dz^{M} b_{MN}(z)=  \frac 1 2 e^B e^Ab_{AB}(z)\; .
$$
Using an analogy with worldvolume vector gauge fields in the Green-Schwarz-like formulations of the D-branes \cite{Cederwall:1996ri,Aganagic:1996pe,Bergshoeff:1996tu}, the authors of \cite{Howe:1996yn} assumed that the field strength of $b_2$ has the following form \footnote{This assumption was later shown to be in agreement \cite{Howe:1997vn,Bandos:1997rq} with the chiral form field strengths appearing in the M5-brane action \cite{Bandos:1997ui,Aganagic:1997zq}.}
\begin{eqnarray}\label{M5:H3=db-=}
H_3:=db_2 - {C}_3 = {1\over 3!}e^c e^b e^a H_{abc}
\; ,   \qquad\end{eqnarray}
It is constrained to have non-zero components only along the bosonic directions of ${\cal M}^{(6|16)}_{sw}$, which is a natural assumption prompted by the fact that $h_{abc}$ carries only the vector indices.

In \eqref{M5:H3=db-=} ${C}_3$ is the pull-back on ${\cal M}^{(6|16)}_{sw}$  of the three-form gauge superfield of 11D supergravity. The explicit form of ${C}_3$ in the flat 11D superspace is somewhat cumbersome, but it is not required for our discussion. We only need to know the form of its field strength $\mathcal F_4$, which is much simpler and manifestly supersymmetric
\begin{eqnarray}
\label{cF4=}
\mathcal F_4=dC_3={1\over 4} E^{\underline{b}}   E^{\underline{a}}   E^{\underline{\alpha}}  E^{\underline{\beta}}\;
 {\Gamma}_{\underline{a}\underline{b}\, \underline{\alpha}\underline{\beta}} \; . \qquad
\end{eqnarray}
To relate $H_3$ to the self-dual tensor $h_{abc}$, we should study the  Bianchi identity
\begin{eqnarray}
\label{M5:dH3=-F4}
dH_3=-{\cal F}_4=-{1\over 4} E^{\underline{b}}  E^{\underline{a}}   E^{\underline{\alpha}}  E^{\underline{\beta}} {\Gamma}_{\underline{a}\underline{b}\, \underline{\alpha}\underline{\beta}} \;\;.
\qquad
\end{eqnarray}
Due to the constraint \eqref{M5:H3=db-=}  on $H_3$,
 the left-hand side of this identity is
\begin{equation}\label{dH3}
dH_3=\frac 12  T^c\,e^b e^a H_{abc}+{1\over 3!}e^c e^b e^a \mathcal D H_{abc}\,,
\end{equation}
where the superworldvolume torsion two-form $T^c$  was defined in  \eqref{dea1}. Substituting \eqref{Ef=eV(h)+} into \eqref{dea1} we get
\begin{eqnarray}\label{M5:Dea=}
 T^a=\mathcal De^a =- 2e^{\alpha q}\wedge e^{\beta p} C_{qp} \gamma^b_{\alpha\beta}\, (\delta_b{}^a-k_b{}^a)\; + 2 e^{b}\wedge e^{\alpha q} C_{qp} (\psi_{b}{}^p\tilde{\gamma}{}^ah)_\alpha + 2e^c\wedge e^b \psi_b^q \tilde{\gamma}{}^a\psi_c^p C_{qp} \; .  \qquad
\nonumber
\end{eqnarray}

If we now equate the $e^{\alpha q}e^{\beta r}e^b e^a$ component of \eqref{dH3} (which appears only in its first term) with the corresponding component of the right-hand side of \eqref{M5:dH3=-F4}, then, taking into account the explicit form of the pullbacks of the supervielbeins $E^{\underline \alpha}$ and $E^{\underline a}$ given in \eqref{pbvs1} and \eqref{Ef=eV(h)+}, and relations between the Lorentz harmonics given in Appendix B, one can show \cite{Howe:1996yn,Howe:1997fb} that $H_{abc}$ and $h_{abc}$ are related to each other in the following way
\begin{eqnarray}
\label{mH=h}
m_a{}^dH_{bcd}=h_{abc},
\end{eqnarray}
or
\begin{eqnarray}
\label{mH=h1}
H_{acd}=m_a^{-1d}h_{dbc}\,,
\end{eqnarray}
where
\begin{eqnarray}\label{M5:mab=}
m_a{}^b=\delta_a{}^b{{-}} k_a{}^b= \delta_a{}^b{{+}} 2h_{acd}h^{bcd}
\end{eqnarray}
and $m_a^{-1b}$ is its inverse
\begin{equation}\label{m-1}
m^{-1}{}_a{}^{b}=\frac 1{1-\frac 19 k_c{}^dk_d{}^c}\left(\delta_a^b{{+}}k_a{}^b\right)\,.
\end{equation}

It can be then shown that the $e^ae^be^ce^d$ component of \eqref{M5:dH3=-F4} produces the second order equation of motion of the chiral gauge field $b_2$
\begin{eqnarray}
\label{M5:DbH=}
\mathcal D_{[a}H_{bcd]} &=& 3!\, C_{qp}(\psi_{[a}\tilde{\gamma}^e\psi_{b})H_{cd]e} \; .
\qquad
\end{eqnarray}
 Note that this equation is actually the Bianchi identity for \eqref{M5:H3=db-=}.

Now, as in the case of the M2-brane, one can proceed with the analysis of the consequences of the superembedding conditions and derive the equations of motion of the M5-brane coordinate functions $\Theta^{\alpha}$ and $X^{\underline a}(z)$ which have the following form (in flat $D=11$ superspace)
\begin{equation}\label{fm5}
m^{ba}(\nabla_a\Theta^{\underline\alpha})(E^{\underline a}_b\Gamma_{\underline a})_{\underline \alpha\underline \beta}v^{\alpha\,\underline\alpha}_q=0\,,
\end{equation}
\begin{equation}\label{bm5}
m^{bc}m_c{}^a(\nabla_aE_b^{\underline a})u^i_{\underline a}=0\,.
\end{equation}
For further details of the derivation of the M5-brane equations from superembedding see the original articles \cite{Howe:1996yn,Howe:1997fb} and the review \cite{Sorokin:1999jx}, and for their equivalence  to the equations obtained from the M5-brane action see \cite{Bandos:1997gm,Sorokin:1999jx}.

\section{Conclusion}
In this Chapter we have described main features of the
superembedding approach, which has not only allowed one to explain
and clarify various classical and quantum properties of
superstring and superbrane theory, but has also found applications
in the construction and description of new superbrane models, and
of field theories with partially broken supersymmetry, as well as
for solving practical problems. For instance, it has been used to derive new solutions of M5-brane equations \cite{Howe:1997ue,Howe:1997hx}, for
calculating vertex operators in (M2-M5)-brane systems
\cite{Moore:2000fs}, constructing higher order contributions
to brane actions \cite{Ivanov:1991ub,Howe:2001wc}, and the study of branes ending on branes \cite{Chu:1997iw,Chu:1998jva,Chu:1998jva,Chu:1998pb}.
 It was also used in search for hypothetical branes \cite{Bandos:2008bn} and the description of exotic branes such as the heterotic 5-brane \cite{Bandos:2011wn},
for the derivation of the equations of motion of multiple $Dp$-brane systems  \cite{Drummond:2002kg,Bandos:2009yp} and  of a multiple M-wave (multiple M0-brane) system  \cite{Bandos:2009gk,Bandos:2010jn,Bandos:2010hc}.
In \cite{Howe:2005jz,Howe:2006rv,Howe:2007eb} a generalization of the embedding approach
to worldvolume superspaces with so-called boundary fermions was developed and used as the basis for the construction of a non-conventional action for
multiple D-branes.

One may expect the superembedding methods be also useful for other purposes, such as a unified (S--duality) description of
 fundamental and solitonic extended objects \cite{Bandos:2000hw} and in the context of Double Field Theory and Exceptional Field Theories.


\section*{Acknowledgements}
\label{secAck}
This work was supported in part by Spainish MCIN/AEI and FEDER(ERDF EU) grant  PID2021-125700NB-C21, and by the Basque Government grant IT-1628-22.

\appendix
\section{$SO(1,2)\times SO(8)$ invariant representation for $D=11$ Dirac matrices and M2-brane Lorentz harmonics}

We use the following $SO(1,2)\times SO(8)$ invariant
representations for 11D Dirac matrices and charge conjugation matrix
\begin{eqnarray}\label{11DG=3+8}
& (\Gamma^{\underline{a}})_{\underline\alpha}{}^{\underline{\beta}}
 \equiv \left(\Gamma^{a}, \Gamma ^{i} \right)\; , \qquad a=0,1,2 \; , \qquad i=1,\ldots, 8 \; , \qquad
\nonumber \\  \label{11DG=3} & (\Gamma ^{a})_{\underline\alpha}{}^{\underline{\beta}}
\equiv \left(\Gamma ^{\underline{ 0}},~\Gamma^{\underline{ 9}}, {}~\Gamma^{\underline{
10}} \right) \equiv \left(\Gamma ^{{ 0}},\Gamma ^{{ 1}},\Gamma ^{{ 2}}\right) = \left(
\begin{matrix}\gamma ^{a~ \beta}_{~\alpha} \delta _{qp} & 0 \cr 0 &
\gamma^{a}{}_{\beta}{}^{\alpha} \delta _{\dot{q} \dot{p}} \end{matrix}\right), \nonumber \\ \label{11DG=8}
& (\Gamma ^{i})_{\underline\alpha}{}^{\underline{\beta}} \equiv \left(\Gamma
^{1},\ldots , \Gamma ^{8}\right)= \left( \begin{matrix} 0 & -i\epsilon _{\alpha \beta} \gamma
^{i}_{{q}\dot{p}} \cr - i \epsilon ^{\alpha \beta} \tilde{\gamma}^{i}_{{\dot q}{p}} & 0
\cr \end{matrix} \right) \nonumber \\
\label{11DC=3+8}
& C^{\underline{\alpha}\underline{\beta}}= - C^{\underline{\beta}\underline{\alpha}}=
{\it diag} \left(i\epsilon ^{\alpha \beta}\delta_{qp},~ i \epsilon_{\alpha \beta}
\delta_{\dot{q}\dot{p}}\right), \quad C_{\underline{\alpha}\underline{\beta}} = \hbox{ {\it diag}} \left( -i \epsilon
_{\alpha \beta} \delta _{qp} ,~ -i \epsilon ^{\alpha\beta} \delta_{\dot{q}\dot{p}}
\right)
\; . \quad
\end{eqnarray}
Here $\gamma^a{}_\alpha{}^{\beta}$ are $SO(1,2)$ Dirac matrices
and $\gamma ^{i}_{q\dot{q}}$ are $SO(8)$
Pauli-like matrices (Klebsh-Gordan coefficients), which obey  the following relations
\begin{eqnarray}
 \label{3Dgammas}
\gamma^a_{\alpha\beta}:= - i \gamma^a{}_\alpha{}^\gamma \epsilon_{\gamma\beta} =
\gamma^a_{\beta\alpha}=\gamma^a_{(\alpha\beta)}\; , \quad {\gamma}_a^{\alpha\beta}:= i \epsilon^{\alpha\gamma}  \gamma^a{}_\gamma{}^\beta= {\gamma}_a^{(\alpha\beta)} \; , \quad \epsilon^{\alpha\gamma}\epsilon_{\gamma\beta}=
\delta^{\alpha}_{\beta}\; , \qquad  \nonumber \\
\gamma^{ab}=-i\epsilon^{abc}\gamma_c \; , \quad
\gamma^a_{\alpha\beta}{\gamma}_a^{\gamma\delta}= 2 \delta_{(\alpha}{}^{\gamma}
\delta_{\beta )}{}^{\delta}
\; , \qquad \\
 \label{8Dgammas}
\tilde{\gamma}^i_{\dot{p} q}:= \gamma^i_{q\dot p} \; , \qquad \gamma^i_{q\dot
p}\gamma^j_{q \dot p} + \gamma^j_{q\dot p}\gamma^i_{q \dot p}= 2\delta^{ij}\delta_{qp}
\; , \qquad \gamma^i_{p\dot q}\gamma^j_{p \dot p} + \gamma^j_{p\dot q}\gamma^i_{p \dot
p}=
2\delta^{ij}\delta_{\dot{q}\dot{p}} \; , \qquad  \nonumber \\
\gamma^i_{q\dot q}\gamma^i_{p \dot p} = \delta_{qp} \delta_{\dot{q}\dot{p}}+ {1\over
4} \gamma^{ij}_{qp} \tilde{\gamma}^{ij}_{\dot{q}\dot{p}}  \qquad \Rightarrow  \qquad
\gamma^i_{(q|\dot q}\gamma^i_{|p) \dot p} = \delta_{qp} \delta_{\dot{q}\dot{p}} =
\gamma^i_{q(\dot q|}\gamma^i_{p |\dot p )}\; . \qquad
\end{eqnarray}
Both 11D and 3d Dirac matrices are imaginary in our mostly minus signature conventions
\begin{eqnarray}\label{11Deta=+--}
\eta^{\underline{a}\underline{b}}=diag(+,-,\ldots ,-)
\; , \qquad \eta^{{a}{b}}=diag(+,-,-)
\; . \qquad
\end{eqnarray}
Using the above representation the relations (\ref{VGVT=uHa}) and (\ref{VTGV=uHa}) betweein the Lorentz harmonic variables adapted to the embedding of the M2-brane worldvolume superspace are
\begin{eqnarray}
 \label{11DuaG=AP}
 u^{~a}_{\underline{b}} \Gamma^{\underline{b}}_{\underline{\alpha}\underline{\beta}} =
v_{\underline{\alpha}}^{\; \alpha q} (\gamma_a )_{\alpha \beta}
v_{\underline{\beta}}^{\; \beta q} + v_{\underline{\alpha} \alpha
\dot{q}} (\gamma_a )^{\alpha \beta} v_{\underline{\beta} \beta
\dot{q}} \; , \qquad \\
 \label{11DuiG=AP}
u^{~i}_{\underline{b}} \Gamma^{\underline{
b}}_{\underline{\alpha}\underline{\beta}} = -2
v_{(\underline{\alpha}|}{}^{\alpha q} \gamma^i_{q \dot{q}}
v_{|\underline{\beta})\; \alpha \dot{q}} \; . \qquad
\end{eqnarray}
The $SO(1,10)$ spinor indices $\underline\alpha$ and $\underline \beta$ are raised and lowered by the charge conjugation matrices \eqref{11DG=3+8}.

The unity decomposition in terms of the Lorentz spinor harmonics is
\begin{eqnarray}\label{I=vv+M2}
\delta_{\underline{\beta}}{}^{\underline{\alpha}} = v_{\underline{\beta}}{}^{\alpha q} v_{\alpha
q}{}^{\underline{\alpha}} +  v_{\underline{\beta} \alpha \dot q}
v^{\underline{\alpha} \alpha \dot q} \; ,\qquad
 \qquad
\end{eqnarray}
while
\begin{eqnarray}\label{bG:=M2}
v_{\underline{\beta} \alpha \dot q}
v^{\underline{\alpha} \alpha \dot q} - v_{\underline{\beta}}{}^{\alpha q} v_{\alpha
q}{}^{\underline{\alpha}} &=& {i\over
3!} \varepsilon_{abc} (v_{\alpha q}{}^{\underline{\alpha}}
\tilde{\gamma}^{abc \, \alpha \beta} v_{\beta
q}{}^{\underline{\gamma}} +
 v^{ \alpha \dot{q}\, \underline{\alpha}} \gamma^{abc}_{\alpha \beta}
v^{ \beta \dot{q}\, \underline{\gamma}})
C_{\underline{\gamma}\underline{\beta}}\nonumber \\
&=&{\bar{\Gamma}}_{\underline{\beta}}{}^{\underline{\alpha}}:= {\frac i{3!}} \varepsilon_{abc}(\Gamma_{\underline a\underline b\underline c} u_{\underline a}{}^au_{\underline b}{}^bu_{\underline c}{}^c)_{\underline\beta}{}^{\underline \alpha}\,.
\end{eqnarray}
Equation  (\ref{DV=VOm})  in the case of the M2-brane splits into
\begin{eqnarray}\label{Dv=M2}
\mathcal Dv_{\underline\beta}^{\alpha q}= {1\over 2} \,\Omega_a{}^{i}\, \gamma^a_{\alpha\beta}
 \gamma^i_{q\dot{p}}  v_{\underline\beta\,\alpha \dot p} \; , \qquad \mathcal Dv_{\underline\beta\,\alpha \dot q}=- {1\over 2} v_{\underline\beta}^{\beta p} \,\Omega_a{}^{i}\, \gamma^a_{\alpha\beta}
 \gamma^i_{p\dot{q}}\; .
\end{eqnarray}

\section{ $SO(1,5)\times SO(5)$ invariant representation for $D=11$ Dirac matrices and  M5-brane Lorentz harmonis  }

The following $SO(1,5)\times SO(5)$ invariant
representations for the 11D Dirac matrices and the charge conjugation matrix are suitable for the description of the M5-brane in the superembedding approach
\begin{eqnarray}\label{11DG=6+5}
& (\Gamma^{\underline{a}})_{\underline\alpha}{}^{\underline{\beta}}
 \equiv \left(\Gamma^{a}, \Gamma ^{i} \right)\; , \qquad a=0,1,\ldots,5 \; , \qquad i=1,\ldots, 5 \; , \qquad
\nonumber
\\
 \label{11DG=6} & (\Gamma ^{a})_{\underline\alpha}{}^{\underline{\beta}}
 = \left( \begin{matrix} 0 & -i\gamma^a_{\alpha \beta} \delta_q{}^p
 \cr i\tilde{\gamma}{}^{a\, \alpha \beta} \delta_q{}^p & 0
\cr  \end{matrix} \right)\;  , \nonumber
 \\
  \label{11DG=5}
& (\Gamma ^{i})_{\underline\alpha}{}^{\underline{\beta}} \equiv \left(\Gamma
^{1},\ldots , \Gamma ^{8}\right)= \left(
\begin{matrix}(\gamma ^iC)_q{}^p \delta_{\alpha}{}^{\beta} & 0 \cr 0 &  -(\gamma ^iC)_q{}^p \delta^{\alpha}{}_{\beta} \end{matrix}\right)
 \nonumber \\
\label{11DC=6+5}
& C^{\underline{\alpha}\underline{\beta}}= - C^{\underline{\beta}\underline{\alpha}}=
 \left( \begin{matrix} 0 & -i\delta^{\alpha}{}_{\beta} C
^{{q}{p}} \cr -i\delta_{\alpha}{}^{\beta} C
^{{q}{p}} & 0 &
\cr \end{matrix} \right)  \; , \quad C_{\underline{\alpha}\underline{\beta}}=
 \left( \begin{matrix} 0 & i\delta_{\alpha}{}^{\beta}  C
_{{q}{p}} \cr i\delta^{\alpha}{}_{\beta} C
_{{q}{p}} & 0
\cr\end{matrix} \right)  \;  . \quad
\end{eqnarray}
The ${\tt Spin}(1,5)\sim SU^*(4) $ Klebsh-Gordan coefficients (Pauli-like matrices) are antisymmetric $4\times 4$ matrices
\begin{equation}\label{4gamma}
\gamma^{a}_{\alpha\beta}=-\gamma^{a}_{\beta\alpha}=\gamma^{a}_{[\alpha\beta]}, \qquad \tilde{\gamma}_{a}^{\alpha\beta}=-\tilde{\gamma}_{a}^{\beta\alpha}=\tilde{\gamma}_{a}^{[\alpha\beta]}\,, \qquad \alpha=1,2,3,4.
\end{equation}
They have the following properties
\begin{eqnarray}
 \label{6Dgammas}
(\gamma^{(a}\tilde{\gamma}{}^{b )})_{\alpha}{}^{\beta} = \eta^{ab}\delta_{\alpha}{}^{\beta}\; , \qquad \eta^{ab}= diag (+,-,-,-,-,-)\; , \qquad \tilde{\gamma}_{a}^{\alpha\beta} ={1\over 2}\epsilon^{\alpha\beta\gamma\delta}{\gamma}_{a \gamma\delta}
\nonumber \\  \gamma^a_{\alpha\beta}\tilde{\gamma}_a^{\gamma\delta}= -4 \delta_{[\alpha}{}^{\gamma}
\delta_{\beta ]}{}^{\delta} \; , \qquad \gamma^a_{\alpha\beta} {\gamma}_{a \gamma\delta}= -2 \epsilon_{\alpha\beta\gamma\delta} \; , \quad  \gamma^{abcdef}{}_{\alpha}{}^{\beta}= \epsilon^{abcdef}\delta_{\alpha}{}^{\beta}  \nonumber \\
\gamma^{abc}{}_{\alpha\beta}=\gamma^{abc}{}_{(\alpha\beta )}= - {1\over 3!}\epsilon^{abcdef}\gamma_{def\;\alpha\beta}
\; , \qquad \tilde{\gamma}^{abc \alpha\beta}=\tilde{\gamma}^{abc (\alpha\beta )}=  {1\over 3!}\epsilon^{abcdef}\gamma_{def}^{\alpha\beta}
\; . \qquad
\end{eqnarray}
The matrices \eqref{4gamma} are pseudo-real in the sense that the conjugate matrices $\gamma^{a *}_{\dot{\alpha}\dot{\beta}}:= (\gamma^{a}_{{\alpha}{\beta}})^*$ are expressed through $\gamma^{a}_{{\alpha}{\beta}}$ with the use of a matrix $B_\alpha{}^{\dot{\alpha}}$ \cite{Howe:1983fr} obeying $BB^*=-I$,
\begin{eqnarray}
 \label{6dG*=BGB}
(B\gamma^{a *}B^T):= B_\alpha{}^{\dot{\alpha}}
\gamma^{a *}_{\dot{\alpha}\dot{\beta}} B_\beta{}^{\dot{\beta}}= \gamma^{a}_{{\alpha}{\beta}}
\; , \qquad (B^{*T}\tilde{\gamma}^{a *}B^{*}):= B^*{}_{\dot{\alpha}}{}^\alpha
\tilde{\gamma}{}^{a * \dot{\alpha}\dot{\beta}} B_\beta{}^{\dot{\beta}}=
\gamma^{a}_{{\alpha}{\beta}}
\; , \qquad \\ \nonumber
B_\alpha{}^{\dot{\beta}}B^*{}_{\dot{\beta}}{}^{\beta}= - \delta_\alpha{}^{\beta}\; . \qquad
\end{eqnarray}

The properties of the ${\tt Spin}(5)\sim USp(4)$ Klebsh-Gordan coefficients (Pauli-like matrices) $\gamma^i_{qr}$ and $\tilde\gamma^{i\,qr}$, and the charge conjugation matrices $C_{qr}$ and $C^{qr}$ ($i,j=1,\ldots, 5$, ${}q,p,r,s=1,\ldots , 4$) are
\begin{eqnarray}\label{Cliff5d} {\gamma}^{i}\tilde{\gamma}^{j}+
{\gamma}^{j}\tilde{{\gamma}}^{i}= 2{\delta}^{ij} \delta_q{}^p \; ,
\qquad {\gamma}^{i}\tilde{\gamma}^{j}- {\gamma}^{j}\tilde{{\gamma}}^{i}=:
2{{\gamma}}^{ij}{}_q{}^p\;  ,  \qquad \\ \label{sigma5d=} {\gamma}^{i}_{qp}=-
{\gamma}^{i}_{pq}= - (\tilde{\gamma}^{i qp})^*=  {1\over 2}\epsilon_{qprs}
\tilde{{\gamma}}^{i\, rs} \; ,  \qquad  \\
\label{5d=C} C_{qp}=-
C_{pq}= - (C^{qp})^* =  {1\over 2}\epsilon_{qprs}
C^{rs} \; ,  \qquad  C_{qr}C^{rp}=\delta_q{}^p\; , \qquad \nonumber \\   C\tilde{\gamma}^iC = - {\gamma}^i \; ,  \qquad C{\gamma}^iC = - \tilde{\gamma}^i \; , \qquad   \\
\label{so(5)id} {\gamma}^{i}_{qp}\tilde{\gamma}^{i rs}= -4
\delta_{[q}{}^{r}\delta_{p]}{}^{s}- C_{qp}C^{rs}\; , \qquad {\gamma}^{i}_{qp}\,
{\gamma}^{i}_{rs} = -2\epsilon_{qprs} - C_{qp}C_{rs} \; .
\end{eqnarray}

The relations (\ref{VGVT=uHa}) and (\ref{VTGV=uHa}) between the Lorentz harmonic variables adapted to the embedding of the M5-brane are
\begin{eqnarray}
 \label{M5:uaG=}
 u^{~a}_{\underline{b}} \Gamma^{\underline{b}}_{\underline{\alpha}\underline{\beta}} =
v_{\underline{\alpha}}^{\; \alpha q} (\gamma_a )_{\alpha \beta}
v_{\underline{\beta}}^{\; \beta p} C_{qp}- v_{\underline{\alpha} \alpha
}{}^q \tilde{\gamma}_a^{\alpha \beta} v_{\underline{\beta} \beta}{}^p C_{qp} \; , \qquad \\
 \label{M5:uiG=}
u^{~i}_{\underline{b}} \Gamma^{\underline{
b}}_{\underline{\alpha}\underline{\beta}} = 2i
v_{(\underline{\alpha}|}{}^{\alpha q} \gamma^i_{q p}
v_{|\underline{\beta})\alpha}{}^p \; , \qquad
\end{eqnarray}

The worldvolume covariant derivatives of the M5-brane Lorentz spinor harmonics have the following form
\begin{eqnarray}\label{Dvaq=}
& \mathcal Dv_{\underline{\alpha}}{}^{\alpha q}= {i\over 2}\Omega^{ai}  v_{\underline{\alpha} \beta}{}^p
\tilde{\gamma}_a^{\beta\alpha} ({\gamma}^iC)_p{}^q \; , \qquad &\mathcal  Dv_{\alpha q}{}^{\underline{\alpha}}= {i\over 2}\Omega^{ai}
{\gamma}_{a \, \alpha\beta} ({\gamma}^iC)_q{}^p v_p^{\beta\underline{\alpha}}\; , \qquad
 \qquad \\ \label{Dvaaq=} &\mathcal  Dv_{\underline{\alpha}\alpha }{}^{q}= - {i\over 2}\Omega^{ai}  v_{\underline{\alpha}}^{\beta p}
\tilde{\gamma}_{a\, \beta\alpha} ({\gamma}^iC)_p{}^q \; , \qquad &\mathcal  Dv_q{}^{\alpha\underline{\alpha}}= - {i\over 2}\Omega^{ai}\tilde{\gamma}_a^{\alpha\beta}  ({\gamma}^iC)_q{}^p v_{\beta p}{}^{\underline{\alpha}}\; . \qquad
 \qquad
\end{eqnarray}


\providecommand{\href}[2]{#2}\begingroup\raggedright\endgroup

\end{document}